\documentclass[prc,showpacs,aps,amssymb,floatfix,endfloats]{revtex4}
\usepackage{psfig}

\def\momi{moment of inertia }
\def\momis{moments of inertia }
\def\am{angular momentum }
\def\J{{\cal J}}
\def\pot{Periodic Orbit Theory }
\def\beq{\begin{equation}}
\def\eeq{\end{equation}}
\def\bea{\begin{eqnarray}}
\def\eea{\end{eqnarray}}
\def\rig{rigid-body }
\def\f{\varphi}
\begin{document}
\bibliographystyle{prsty}
% \setlength{\parindent}{1.5em}
%\draft

\title{Gross shell structure at high spin in heavy nuclei}
\author{
M.A.\,Deleplanque$^1$,
S.\,Frauendorf$^2$,
V.V.\,Pashkevich$^3$,
S.Y.\,Chu$^1$,
and
A.\,Unzhakova,$^3$}
\affiliation{$^1$ Nuclear Science Division, 
Lawrence Berkeley National Laboratory, 
         Berkeley, California 94720}
\affiliation{$^2$ Physics Department, University of Notre Dame,
	 Notre Dame, Indiana 46556}
\affiliation{$^3$ Joint Institute of Nuclear Research, 
         Dubna, Russia}
\date{\today}
\begin{abstract}
Experimental nuclear moments of inertia at high spins along the yrast line 
have been determined systematically and found to differ from the rigid-body values.  
The difference is attributed to shell effects and these have been 
calculated microscopically.  The data and quantal calculations are interpreted 
by means of the semiclassical Periodic Orbit Theory.  From this new perspective, 
features in the moments of inertia as a function of neutron number and spin, as well 
as their relation to the shell energies can be understood.  Gross shell effects 
persist up to the highest angular momenta observed.
\end{abstract}
%\vspace*{9cm}
%
%\newpage
%
% ftp://aps.org/pub/pacs/pacs_99.asc
%
% 21.10.Re  Collective levels
% 21.60.Ev  Collective models
% 23.20.Lv  Gamma transitions and level energies
\maketitle

\section{Introduction}

Microscopic calculations based on the rotating mean field (self-consistent
cranking model) describe the rotational energies well.
Although there are numerous detailed comparisons between experiment 
and calculations in the literature, a qualitative understanding of the
rotational response has not yet been reached.  Two idealized 
models have been studied in
detail: the Fermi gas and the harmonic oscillator \cite{bom75}. The Fermi gas
does not take into account the shell structure. Its yrast line
 is characterized by $\J_{rig}$,
the moment of inertia of a rigidly rotating mass distribution (the rigid-body moment of inertia). 
The harmonic oscillator has a very specific shell
structure, which is different from the one of the nucleus. At moderate
angular momentum and equilibrium deformation it has also the rigid-body
moment of inertia.  However, we know from experiment that the 
moment of inertia along the yrast line is in general different from 
${\cal J}_{rig}$, indicating that there are currents in the rotating 
frame. The  moments of inertia observed at low spin 
reach only about 40\% 
of the rigid-body value. This  reduction has been commonly attributed to the 
presence of pair correlations \cite{bom75}. If the nucleus were a macroscopic 
superfluid
it  would have the irrotational flow pattern, which corresponds  to strong 
counter currents in the rotational frame and 
\momis $\J_{irrot}$ that are substantially smaller than $\J_{rig}$.
The observed \momis lie in between $\J_{irrot}$ and $\J_{rig}$. 
 This can be explained by the fact
that the coherence length of nuclear pairing is larger than the nuclear size,
which prevents the nucleus from fully developing the irrotational 
flow pattern of a macroscopic superfluid. 

However, the \momis observed
at high spin also substantially deviate from the rigid-body value, although
the pair correlations are expected to be small. This observation
is corroborated by self-consistent cranking
 calculations that assume zero pairing (see e. g. \cite{nir95,afa99,fra00}),
which find strong deviations from $\J_{rig}$ in accordance with experiment.
Nevertheless, the question of long 
standing interest in nuclear physics,  how
is angular momentum generated in nuclei, is not yet answered in a
 satisfactory way.  Shell effects 
(i.e., the bunching of single-particle energy levels) play an
important role   at high angular momentum. 
Superdeformation \cite{jan91}, band termination \cite{afa99}
 and uniform rotation about
a tilted axis \cite{fra01}  are dramatic manifestations of shell effects.  
The shell effects determine the \momis  along the 
yrast line, the importance of which was pointed out by Pashkevich and
Frauendorf \cite{pas74} long ago. The deviations 
from the rigid-body moment of inertia  imply 
that the flow pattern must substantially deviate from the current
of a rigidly rotating mass distribution, i. e. there  are strong
net currents in the frame rotating with the density of the nucleus. 
In the present paper we 
investigate the deviations of the \momis from $\J_{rig}$ in a systematic way and give a 
 qualitative interpretation in the frame of the
semiclassical  Periodic Orbit Theory \cite{bba97}, which 
relates the shell structure to classical periodic 
orbits in the nuclear potential. This approach becomes a very instructive 
tool for interpretation if one is interested in the gross
structure, which is determined by only the few shortest orbits. 
We take a fresh look at the \am generation in
nuclei through this perspective.

In section \ref{s:exp}, we extract the experimental 
\momis at high angular momentum, where the pair correlations
are expected to be negligible or at least drastically reduced and 
calculate their deviations from the \rig value.
We find substantial deviations from the rigid-body moments of 
inertia.       
In section \ref{s:qmc}, 
these experimental deviations are compared with mean-field
calculations, which assume zero pairing and employ the shell-correction 
version of the self-consistent cranking model
 (Strutinsky-type calculations in a rotating
Woods-Saxon potential). We find that the calculated deviations 
from the rigid-body moments of inertia are rather similar to the 
experimental ones. We also extract experimental shell energies which
are compared with the calculated ones. 
In section \ref{s:pot}, we review  the elements of the  \pot which we need
to explain some features of the   gross shell structure at 
high \am.   We find that the shell energies and the shell \momis
 are correlated and determined by the
contributions from the shortest classical periodic 
orbits. The competition and interference between the orbits in the 
meridian planes and
 the equatorial plane of the deformed potential account for the most conspicuous
features of the shell energies and of 
the deviations of the \momis from the \rig 
value. They also explain the appearance  of regions of high-spin 
isomers. 
The deviations of the  \momis from the rigid-body
value indicate the presence of currents in addition to the simple
flow pattern of a rigidly rotating mass distribution, i.e., of net
currents in the rotating body-fixed frame of reference.
A first  report of our results was given in ref. \cite{ber02}.

\section{Extraction of  moments of inertia and shell energies from the  data}
\label{s:exp}

We are interested in the {\em gross shell structure}, not in the details of
the dependence of the energy on the particle numbers $Z$ and $N$ and the 
\am $I$. Accordingly, we characterize the yrast line (sequence
of states with minimal energy at a given $I$) by only one parameter,
the moment of inertia $\J$, assuming for the energy the $I$-dependence,
\beq\label{eyrast}
E(I)=E_o+\frac{I(I+1)}{2\J}.   
\eeq
The moment of inertia $\J$ we want to determine for each nucleus is 
the {\it yrast\/} moment of inertia in the  terminology 
introduced by Bohr and Mottelson (see Fig.1 in \cite{bom81}).  
It characterizes the increase in energy of the yrast levels with 
spin $I$ and represents the global characteristics of the generation of 
angular momentum in the nucleus.
It should be emphasized that the yrast line does 
not necessarily coincide with one rotational band. It may be composed of
pieces of crossing bands or not contain bands at all.
Since the expression (\ref{eyrast}) describes well the 
average $I$ dependence of the yrast at high spin (see below), there is no need
to distinguish between {\it kinematical} and {\it dynamical} 
moments of inertia.
  In Fig.~\ref{fig:yrast}, examples of the 
yrast energies as  functions of $I(I+1)$ are given for a 
"rotational" nucleus (top, $^{160}Er$) and for a "non-rotational" 
nucleus (bottom, $^{150}Dy$).  While the plot is smooth for the 
rotational nucleus, due to the presence of long rotational bands, 
there are irregularities for the non-rotational nucleus,
 which reflect  alignments of individual nucleons with different angular momentum.  
Besides these fluctuations, there is an approximate linear relationship 
 at high spin in both cases. Since we are interested  in the
{\it average\/} behavior along 
the yrast line, we fit the linear relation (\ref{eyrast}) to
 the {\it yrast\/} line 
at the highest spins, which will give the {\it yrast\/} moment of 
inertia. We do not fit separate bands, which would give a 
{\it band} moment of inertia.  This procedure is applied to the rotational 
nuclei and to the non-rotational 
nuclei for which the yrast line is  irregular. 

\subsection{Quenching of the pair correlations at high spin}

Generating  \am  destroys the pair correlations. This
is in analogy to the transition from the paired to the 
normal phase of a superconductor  in a magnetic field. In a 
finite nucleus, the transition is less distinct and stretched over
a substantial interval of angular momentum. Hence, there is a question as to what
extent one can 
extract from the high-spin data \momis that characterize the rotation
of the unpaired nucleus.  The quenching of pairing by rotation
has been studied in a number of publications. See e.g. the recent studies
\cite{shi89,shi93,shi95,alm01},
which refer to the extensive earlier work. The following picture emerges.
The typical pairing effects, which include the large reduction of the 
\momi at low spin, are caused by strong correlations  in the 
occupation of a few levels near the Fermi
surface. They are well accounted for by the mean-field 
(Hartree-Fock-Bogoljubov) description of pairing. These so-called
static pair correlations are destroyed by breaking few  pairs. 
The destruction of the static correlations is seen as a change of
the slope of the yrast sequence, which typically appears in the range
$14<I<26$ for the considered nuclei. On top of the static pairing there
are the  dynamic pair correlations, which involve the levels far from 
the Fermi surface. They decrease slowly with increasing angular 
momentum, remaining substantial
in the experimentally accessible spin range. They reduce the \momis
by about 5\%, where the reduction is not sensitive to the detailed
structure of the states near the Fermi surface. This explains why
unpaired cranking calculations (see e.g.  \cite{afa99}) 
are so successful in describing the level structure in the yrast region
 at high spin. 
Hence, we try to extract the \momis from the region where the
static pairing is destroyed. These values are expected to be close
to the \momis of the unpaired nucleus, with a slight overall reduction.
In particular, the experimental \momis will show  
the shell structure of the unpaired nucleus.

\subsection{The extraction method}

As seen in Fig. \ref{fig:yrast}, 
 the experimental excitation energies $E(I)$ when plotted vs $I(I+1)$
are roughly linear at high spin, where the static pairing is destroyed.  
The experimental curve bends sharply down at low spins ($I<15$ and most 
often $I<10$, especially in lighter nuclei). This is caused by the onset
of the static pair correlations, which reduce the moment of inertia.  
The examples are typical for all 
the studied nuclei.  
 Since we want to eliminate the 
pairing effects, we  fit the yrast lines with a straight 
line for the highest $10 \hbar$ where possible, including both 
odd and even spins, as shown in Fig. \ref{fig:yrast}.
If negative  parity levels enter the yrast line they are accepted.   
To avoid interference from low-spin pairing 
effects, we  include only  nuclei that are known up to 
sufficiently high spins, using the following criteria:  the highest known 
spins should be greater than $15 \hbar$ for nuclei with neutron number $N<50$, 
greater than $16 \hbar$ for $50<N<82$, and greater than $17 \hbar$ for 
$N>82$.  The minimum spins entering the fit should be greater than $7 \hbar$ for 
$N<50$, greater than $10 \hbar$ for $50<N<82$, and greater than $12 \hbar$ for 
$N>82$.  Generally the maximum spins are considerably higher than the limits 
above. 
The experimental yrast energies are taken from the ENSDF files using the 
Isotope Explorer program \cite{chu97}, for nuclei with even Z 
in the mass range from 80 to 220.
From the straight line of the fit, $E(I)_{fit}=E_o+I(I+1)/2\J_{exp}$, 
we derive the experimental \momi $\J_{exp}$  and the ground-state energy
$E_o$ of the unpaired nucleus. 
Fig. \ref{fig:pairen} shows the difference $E(I)-E_{fit}(I)$ for a
 representative selection of nuclei. At high spin, the curves
 largely flatten out near
the zero line. There is a distinct drop of the 
yrast levels with decreasing $I$, which is interpreted as
the gain in energy caused by the static pair correlations. 
The onset of this correlations is seen both in rotational and
spherical nuclei. The difference $E_o-E(0)$ may be interpreted as the
 experimental ground state pair correlation energy. It turns out to be 
rather different for the examples. It would be interesting to see if 
theory can reproduce these variations in a systematic way.   
The deviation of the upper portion of $E(I)-E_{fit}(I)$ from the zero line
reflects the uncertainty of the experimental moments of inertia.

The rigid-body moment of inertia is calculated 
using the formula
\begin{equation}
{\cal J}_{rig} [{\hbar}^2 /MeV]= 0.01253 A^{5/3} 
(1+0.5 \sqrt{5/4 \pi} \beta_g) + 0.048225 A, 
\end{equation}  
where $\beta_g$ is the 
ground-state deformation as calculated by M{\"{o}}ller {\it et al.}  
\cite{mon95}.  Since the rigid-body moments of inertia are rather 
insensitive to the deformation, taking the ground-state deformation 
instead of the actual deformation at high spin 
will have only a small effect.  We then calculate 
$ {\cal J}_{yrast} - {\cal J}_{rig}$, the deviation from the 
rigid-body moment of inertia, and plot it as a function 
of neutron number. The experimental moments of inertia, and their 
deviation from rigid-body values for the nuclei with small or  normal
deformation are shown in Figs. \ref{fig:momex} and \ref{fig:dmomex} 
respectively.    
 The superdeformed bands of the $A=150$ region are shown in Fig. 
\ref{fig:dmomexsd}. For $I$, we use the estimates of
 \cite{rag93}.

The shell energies at spin $I$ are calculated 
in the Strutinsky sense \cite{str68} 
by means of  the  formula
\begin{equation}
E_{sh}(I) = E(I)-B - \left(E_{LD} + \frac{I(I+1)}{2\J_{rig}}\right).
\end{equation}
where $B$ is the experimental binding energy and
 $E_{LD}$ is the liquid-drop energy  at the ground-state deformation taken
from \cite{mon95}.  Fig. \ref{fig:deex0} shows the ground-state
shell energies and Fig. \ref{fig:deex20} the shell energies at $I=20$. 
The energy E$_{sh}$ also contains 
the fluctuations of the pair correlation energy since the smooth pair energy 
is included in E$_{LD}$.

As seen in Fig. \ref{fig:dmomex}, the deviations of the \momis 
from the rigid-body value are dramatic for small and normal deformation.
  Rough numbers for the experimental 
deviations are -25\% for $N \sim 75$, +20\% for $N\sim 86$,
 -15\% for $N \sim 95$, 
-40\% for $N \sim 100$, and  -65\% for $N \sim 120$. 
Note the positive value, which cannot possibly be attributed to pair 
correlations. 
We are going to interpret the deviations in terms of 
the shell structure. Here, we want to
argue that they are not caused by the limited \am reached in  experiment. 
Fig. \ref{fig:imax} shows
the maximal value of $I$ of each nucleus used in 
the analysis. There is no obvious correlation with Fig. 
\ref{fig:dmomex}, except  near the
closed shells $N=82$ and 126, where both  the \momis and the maximal
values if $I$ are small. In Fig. \ref{fig:dmomex20}, we show
only the nuclides for which the maximum spin is larger than 20.  
They represent approximately 65\% of all nuclei considered.
The overall picture of deviations of the \momis from the rigid-body 
value is very much the same as in Fig. \ref{fig:dmomex}.
Since for the subset of nuclei with the 
highest spins the pairing correlations are, in any case, expected 
to be weaker than at lower spin,
the similarity of the structures in Figs.  \ref{fig:dmomex} and 
\ref{fig:dmomex20} indicates 
that neither the variations in experimental spins nor pairing correlations 
are creating that structure. 
Hence we conclude that Figs. \ref{fig:dmomex} and \ref{fig:dmomex20}
show the contribution of the shell structure to the \momi and are not
related to pairing effects.

\section{Comparison with quantum mechanical calculations}\label{s:qmc}

In this section we demonstrate that the
experimental shell-energies and moments of inertia
can be accounted for by a quantal treatment of the rotating mean field without
pair correlations. 

\subsection{Cranked Woods-Saxon-Strutinky calculations}\label{s:qmcws}

For a realistic quantum mechanical 
calculation we use the Strutinsky-type shell correction
for rotating nuclei \cite{nee76}.
 The single-particle levels are calculated 
at the rotational frequency $\omega$ by diagonalizing 
the cranked deformed Woods-Saxon potential
\beq
h'=t+V(\alpha,\alpha_4)-\vec \omega \cdot \vec j,
\eeq 
where $\vec j$ is the single-particle  \am.
 For the single-particle potential a simple generalization of the
 spherical Woods-Saxon potential is used
 \cite{pas69,dam69}. It is supposed that the potential
 depends on the distance from the surface $l_d(\vec r)$.
 \beq V(\vec r)=V_0/[1+exp(l_d(\vec r)/a)],\eeq
 
  \begin{table}
 \caption{\label{tab:wspar} Parameters of the mean-field
 potential \cite{gar73}.}
\begin{ruledtabular}
 \begin{tabular}{lccccc}
 &$r_0(fm)$&$a(fm)$&$\kappa(fm^{2})$&$V_{mean}(MeV)$&$c^{iso}$\\
 \hline
   Protons & 1.245 & 0.599 & 0.341\\
           &       &       &        & 52.2 & 0.746 \\
  Neutrons & 1.260 & 0.614 & 0.412\\
 \end{tabular}
\end{ruledtabular}
 \end{table}

 \beq  V_{0} =
  - V_{mean} (1 \pm c^{iso}\frac{N-Z}{A}), \eeq
 where the upper (lower) sign refers to protons (neutrons).
 The distance $l_d(\vec r)$ is determined numerically. The sign of
 $l_d(\vec r)$ is taken to be negative inside the nucleus and
 positive outside. In case of the sphere
 $l_d(\vec r)=r_{sph}-R_0$, where $R_0=r_0 A^{\frac{1}{3}}$.
 The spin-orbit interaction is expressed in terms of the central
 potential $V$ as follows
 \beq  V^{so}=-\frac{\kappa}{{\hbar}^2} {\bf \nabla} V \cdot
    [{\bf \sigma p}]. \eeq
 The parameters of the potential are given in table
 \ref{tab:wspar}. They are taken the same for all nuclei considered as the
 averaged of the four sets of the parameters given for the rare earth
 region in ref. \cite{gar73}. The parameter $r_0$ enters in the definition
 of the potential through the volume conservation condition.

 The axially symmetric shape is characterized in the meridian plane
 in the coordinate system in which one of the two families of the
 coordinate lines are the con-focal Cassini ovals \cite{pas71}.
For small deformations the shapes are almost spheroidal
and the 
parameters $\alpha$ and $\alpha_4$ are close to the 
usual quadrupole and hexadecapole deformation parameters $\varepsilon$ and
$\varepsilon_4$.  The shell energy  
(Strutinsky shell correction) $E_{sh}(\alpha, \alpha_4, \omega )$
is obtained from the single-particle levels by the Strutinsky's averaging
procedure \cite{str68}. The deformation parameters are determined 
by minimizing with respect to $\alpha$ and $\alpha_4$ the total 
Routhian (energy in the rotating frame) 
\begin{equation}
E'(\alpha, \alpha_4, \omega) =  E_{sh}(\alpha, \alpha_4, \omega) + 
E_{LD} (\alpha, \alpha_4) - 
\frac{\omega^2}{2}{\cal J}_{rig,\nu}(\alpha, \alpha_4),
\end{equation}
at the given rotational frequency $\omega$.
We consider the two possibilities
that the rotational axis $\vec \omega$
 is perpendicular to the  symmetry axis ($\nu=\bot$)
and that it is parallel ($\nu=\Vert$).
The \am is the expectation value of its projection on the rotational axis,
\beq
J_\nu(\omega)=<\omega|j_\nu|\omega>,
\eeq
where $\nu=\bot$ or $\Vert$ and
 $|\omega>$ is the lowest configuration in the cranked
 Woods-Saxon potential \footnote{
 A Strutinsky renormalization of the \momi does not 
give anything new because the smooth part and the rigid body value of 
$\J$ agree very well (see \cite{pas74}).}. 
 The expression for the liquid-drop energy  $E_{LD}$ is
 given in \cite{pas71}.  The 
calculated moment of inertia at the frequency $\omega$ is 
\begin{equation} 
{\cal J}_\nu(\omega) = \frac{ J_\nu(\omega)}{\omega }.
\end{equation} 
In order   
to emphasize the role of shell effects, we call the deviation of the
moment of inertia 
from the rigid-body value in absence of pairing,
\beq
{\cal J}_{sh,\nu}(\omega)= {\cal J}_\nu(\omega,\alpha, \alpha_4)
-{\cal J}_{rig,\nu}(\alpha, \alpha_4),
\eeq
 the {\em shell} moment of inertia. We calculate the shell \momis
for $\omega=0.3~MeV/\hbar$ in 
the range of nuclei for which there is experimental data.  The nucleus 
is allowed to rotate around the symmetry axis and around an axis 
perpendicular to the symmetry axis, and the shape parameters are
optimized separately for prolate and oblate shape. Figs.~\ref{fig:dmomcafix}
shows the four types of shell \momis.
Out of the four calculations, the mode 
with the lowest total energy is chosen to calculate  the 
optimal shell energies and the optimal
shell moments of inertia. The results are shown in 
Figs. \ref{fig:dmomca} and \ref{fig:deca30}.

\subsection{Discussion}\label{s:qmcd}

The experimental shell  \momis
 are shown in Fig.~\ref{fig:dmomex} as functions
 of neutron number $N$.   The proton numbers 
$Z$ are represented by the different symbols. 
The calculated shell  moments of inertia are 
shown in Fig.~\ref{fig:dmomca}.  In both cases 
the yrast moments of inertia deviate substantially 
from the rigid-body values.  
  The variation of the 
shell \momis with neutron number is 
very similar in the experiment and in the calculation:  there 
are dips at the spherical neutron magic numbers 50, 82 and 126 
and peaks just above and below N=82 then lower values around 
N=90 where the deformation sets in, then another peak around 
N=110 which is the region of high-K isomers.  We will 
see later that these features can be related to particular 
properties of the nucleus (such as deformation, closed shells and axis 
of rotation).  Overall, the experimental shell moments of inertia 
are shifted by about 10\% to the negative side compared to the 
calculated ones.  We attribute this shift to the residual pair
correlations, which are mostly dynamical.

The similarity between the \momis extracted from
the experiment and from the calculations that 
{\it do not include\/} pairing effects  is a confirmation 
that the differences, $\J-{\cal J}_{rig}$,  are not due to pairing effects, 
but are the manifestations of shell effects. In fact, the present
comparison is just another example of the observation that the details of 
the rotational response at high spin are well reproduced by calculations
without pair correlations,  which often give \momis that are very different 
from the rigid-body value. One case is the moments of inertia in 
high-K multi quasiparticle bands \cite{fra00}. The low 
moment of inertia is due to the presence of  orbitals at the Fermi
 surface that  are very  strongly coupled to the 
deformed potential, which makes the generation 
of collective angular momentum costly. Another case is the  smooth 
terminating bands, which typically have a moment of inertia lower than 
${\cal J}_{rig}$ that is well reproduced by calculations without pairing 
\cite{afa99}. The reduction is explained by 
shell effects, i.e., gaps in the single-particle level density for 
certain nucleon numbers.  The angular momentum is generated by 
gradual alignment of the individual angular momenta of the particles 
and holes in incomplete shells, which becomes more and 
more difficult with increasing $I$  causing the decrease in moment of inertia.
Without any doubt, there are many more examples of substantial deviations
of the experimental \momis from the rigid-body value that have been reproduced
by cranking calculations without pair correlations. 
We consider these results as further
 evidence that the deviations are caused by the shell structure and 
not pair correlations. This interesting fact, which has not been 
pointed out enough, will be studied in a systematic way in what follows.

Fig. \ref{fig:deex0} shows the experimental ground-state shell 
energies. They are well reproduced by the calculations in 
Ref. \cite{mon95}
and other mean-field calculations. The $N$ dependence is governed 
by the shell structure of the unpaired single-particle levels.
Both the calculated (Fig. \ref{fig:dmomca}) and experimental (Figs.
\ref{fig:dmomex} and \ref{fig:dmomex20}) shell moments of inertia 
show a structure similar to that of the shell 
energies at zero spin, namely: (1) minima at the same magic numbers; (2) 
higher values just below and above the magic numbers than further 
away; (3) a shoulder or peak around $N = 110$. The shell energies
reflect the level density at the Fermi surface. They are negative if 
the level density is lower than the average and positive if it 
is larger. The relation between the level density and 
the \momi appears directly in 
the statistical estimate for the moment of inertia (cf. \cite{bom75})
$\J=g<l^2>$, where $g$ is the level density  and
$<l^2>$ the average single-particle angular momentum, both
taken near the Fermi surface. The Inglis \momi (see \cite{bom75})
for collective rotation, 
$\sum_{ph}| l_{ph}|^2/e_{ph}$, can be 
estimated by the same expression if we assume that the particle-hole energy
$e_{ph}\sim 1/g$ and $\sum_{ph}| l_{ph}|^2\sim < l^2>$.
Using for $<l^2>$ the classical value and for $g$ the Fermi gas 
value, one obtains the rigid-body value \cite{bom75}. 
The deviations 
from this value are mostly due to the deviations of $g$ from
its smooth (Fermi gas) value, 
i.e. due to the shell structure in the level density.
This just means it is hard to generate \am if the level density is low,
so that the \momi is small; and vice versa. 
Since both the ground-state shell energy and the shell 
\momi are proportional to the level density, their correlation is 
not a surprise.
In section \ref{s:potmomi}, we shall discuss the relationship
between the ground-state shell energies and the shell part of the \momis
in a more quantitative way.
Note that 
in Figs. \ref{fig:deex0}, \ref{fig:deex20} and \ref{fig:deca30}
 the shell energy has been multiplied by a 
factor $A^{4/3}$ where $A$ is the mass of each nucleus.
In section \ref{s:potmomi}, we shall derive this scaling factor, 
which makes the correlation particularly obvious.

The experimental 
shell energies at spin 20 (Fig.~\ref{fig:deex20}) 
look similar to the ones at zero spin.
 However, the
changes with $N$ are less rapid. The
 minima around the main shell closures in 
($N$ = 50, 82 and 126) are still present but are 
less pronounced than in the ground state 
(see Fig.~\ref{fig:deex0}). The same is true for the calculated 
shell energies at $\hbar \omega=0.3~MeV$ (Fig.~\ref{fig:deca30})
as compared to the calculated ones at zero frequency (not shown).
The reason for the damping of the shell structure  
is the correlation between the zero-spin shell 
energy and the moment of inertia, which implies an anticorrelation
between the rotational and the zero-spin shell energy. 
When the zero-spin shell energy is small the rotational energy is large and
vice versa. A quantitative estimate will be given in Section \ref{s:potsefs}.

Concluding this discussion we claim that the deviations of the 
\momis from the rigid-body value at high spin 
are determined by the shell structure
of a system of independent Fermions confined by a leptodermous potential.
This implies analogies with non-nuclear systems of confined Fermions.
Metallic clusters and quantum dots in a magnetic field are examples
(see \cite{fra01} for a review of the analogies),
which we shall refer to in the general discussion of the shell effects
in the next section. The strong deviations of the \momis from
the rigid-body value imply that the flow pattern of the mass current 
must be rather different from the one of a rigidly rotating mass distribution.
In the case of clusters and quantum dots, the difference between the
quantal and the rigid-body currents is the electrical current
carried by the conduction electrons, which gives rise to an unusually large
magnetic susceptibility (see e.g., \cite{fra98}).
Its name ``normal persistent current'' alludes to the
fact that the large magnetic moment is generated  by persistent currents
like in a superconductor (Meissner effect), but that the origin of 
the currents is quite different because the system is in the normal state.  
The analogy  is beautiful. The  nuclear \momis at high spin may be
substantially smaller than the rigid-body value as if the nucleus were 
superfluid. However, it is in the normal state and the reduction is caused by 
the shell effects. 
In the next section we use  the  Periodic Orbit Theory
to obtain more insight into the 
nature of the shell effects and the mass currents that 
occur in the rotating frame at high angular momenta.

\section{Interpretation in terms of the Periodic Orbit Theory}\label{s:pot}

In the preceding section the shell effects in the 
rotational energy were calculated from the quantal levels in the 
rotating potential using Strutinsky's shell correction method
\cite{str68}.  Now, we want to study them from a different 
perspective.  Using the  semiclassical \pot (POT), which does not require 
full quantal calculations, we can understand some global 
characteristics of the rotating many-fermion system.
POT has been successfully used to explain various aspects of shell structure
in fermion systems. The  shell structure of a spherical cavity was first
analyzed by Balian and Bloch \cite{bbl72}. They could relate the 
spacing between the shells to the length of the shortest orbits (the
 triangle and square) and they predicted a long wavelength modulation
 which results from the interferences between these orbits. This supershell
structure was recently found in metal clusters \cite{bjo90}. 
 Strutinsky et al. \cite{str77} first explained how the deformation 
of nuclei is determined by the classical orbits in the meridian planes of
 a spheroidal cavity.
 
\subsection{Basics of the Periodic Orbit Theory}\label{s:potbasics}
A detailed presentation of the \pot (POT)
was given in the book by Brack and Bhaduri
\cite{bba97}. Here we  review some basic facts needed for our discussion.
For  a given potential,  the level density is decomposed into an oscillating
 part, which represents the shell structure, and a smooth background.
The oscillations show up in  related quantities such as the energy and the 
moments of inertia, and they are 
the origin of many structural features of nuclei. 
 POT calculates 
the oscillating part of the level density and of the  derived quantities
in terms of classical periodic orbits in the same potential.
More specifically,  it aims at the oscillating part of the level density
that is averaged over a certain energy interval. For a sufficiently wide 
averaging interval, only the gross structure remains, which is described 
by a few short orbits. This tremendous simplification makes POT a powerful tool 
for interpreting the gross shell structure.  We are interested in the gross 
shell structure of the rotational energy,
which has  been studied before only for the special cases of rotation about the
symmetry axis \cite{kol79} and the harmonic oscillator \cite{bom81,nir95}. 

Let us start with the Fermi gas that describes the smooth behavior without
the shell structure.
We consider $N$ Fermions of one kind
 in a spherical cavity of radius
\beq\label{Ro}
R_o=r_oN^{1/3},
\eeq
two of which occupy one orbital state. 
 It is useful to introduce the wave number, 
$k=\sqrt{2me/\hbar^2}$, where $e$ is the level energy.
The relation
\beq\label{kFRo}
k_FR_o=\left(\frac{9\pi}{4}\right)^{1/3}N^{1/3}\equiv b N^{1/3},~~b=1.92  
\eeq
determines the Fermi wave number $k_F$. Using (\ref{Ro}), we have
$k_Fr_o=b$.
The Fermi   
energy $e_F=(\hbar k_F)^2/2m$ is a convenient energy unit.
The  smooth level density is 
\beq\label{smoothg}
 \tilde g(e)=~\frac{3 N k}{2e_Fk_F}=\frac{3 N e^{1/2}}{2e_F^{3/2}}.
\eeq
The smooth level density in k space is:
\beq
\tilde g(k) = dN/dk = \tilde g(e) (de/dk) =~\frac{3 N k^2}{k_F^3}.
\eeq
The expressions for the Fermi gas
do not depend on the shape of the cavity. They also apply
 to a deformed cavity with the same volume. However, they
describe the smooth behavior only correctly in leading order of $k$. 
The next order (see \cite{bbl72}), 
which accounts for the surface effects, somewhat 
modifies the relations (\ref{kFRo},\ref{smoothg}). These shape dependent
modifications are not important for the qualitative discussions of the 
shell structure in the following.

The level density $g$  
is decomposed into the smooth part $\tilde g$ and an 
oscillating part $g_{sh}$, which contains the shell structure,
\beq
g=\tilde g+g_{sh}.
\eeq
 The oscillating part is given by
\begin{equation}\label{g}
g_{sh}= \sum_{\beta} g_{\beta}
\end{equation}
where $\beta$ labels the periodic orbits that contribute.

The shell structure of the levels in a spherical cavity was studied by
Balian and Bloch \cite{bbl74}.
The classical  system corresponds to a  point mass inside a hollow sphere
bouncing  elastically from the walls.  The  periodic
orbits are  polygons in a plane that contains the center of the
sphere. Fig. \ref{fig:orbits} (top) shows the simplest, the triangle and the square.
 Each polygon generates a family $\beta$,
which consists of all orientations  such a planar
orbit can have within the sphere.
The family is characterized by the number of vertices $v$ (number of
reflections on the surface) and the winding
number $w$ (number of turns around the center until the orbit is
closed), i.e., $\beta=\{v,w\}$. The triangle is \{3,1\}, the square
\{4,1\} and the five-point star \{5,2\} (see Fig. \ref{fig:orbitsd}). 
The explicit contribution from each family is given by
\begin{equation}\label{gbeta}  
g_\beta(k)=  {\cal A}_{\beta}(k) \sin (L_{\beta}k+\nu_{\beta})
D\left(\frac{kL_\beta}{\gamma R_o}\right).
\end{equation}
To be brief, we shall refer to the whole family of such degenerate orbits as
the triangle, square, etc.

As a function of $k$, each term in the sum oscillates with 
the frequency given by the length $L_\beta$ of the orbital
\beq\label{lenght}
L_{v,w}=2v\sin{\f_{v,w}} R_o, ~~~\f_{v,w}=\frac{\pi w}{v},
\eeq
where $\f_{v,w}$ is half the opening angle of one polygon segment.
The Maslov index 
\beq
\nu_\beta =\nu_{v,w}=-\frac{3 v \pi }{2}+\pi w+\frac{3 \pi }{4}
\eeq
is a constant phase, which takes into account that each bounce at the surface
and each turn around the center changes the phase by a constant 
(see \cite{bba97}).   
The amplitude ${\cal A}_{\beta}$ 
depends on the degeneracy of the periodic orbit:  the more 
symmetries a system has, the greater the 
degeneracy, and the more pronounced are the fluctuations of the level 
density. In the case of the spherical cavity,
\beq\label{amp} 
{\cal A}_{v,w}=2b^{5/2}N^{5/6}\frac{k^{3/2}}{k_F^{5/2}}\sin(2
\f_{v,w})\sqrt{\frac{\sin{\f_{v,w}}}{\pi v}}.
\eeq
The relative amplitude of shell oscillations compared to
the smooth level density $\tilde g(e)$ is 
$\sim (N)^{-1/6}$, that is they are comparable for nuclei.   
The amplitude decreases with the opening angle $\f_{v,w}$ of 
the polygon sections, i.e., it decreases 
as the number of vertices $v$ increases. 
Hence the triangle and the square have the
largest amplitudes.

The   damping factor $D(kL_\beta/\gamma R_o)$ is a  decreasing function
of its argument. Its concrete form depends on how the level density is
averaged over $k$. The wider the averaging interval $\gamma$ the more
rapidly long orbits are suppressed. 	In the present paper, we do not
explicitly average. We consider only the rough  
dependence of several quantities on the particle number. 
This gross shell structure can be thought of as the result of averaging
over a fraction of a shell. Hence, it can be understood by 
considering the interplay of the shortest orbits, which are 
the triangle and the square in the case 
of the sphere. Note, we have disregarded the diameter orbit,
although it is the shortest. It gives only a small contribution, 
because it has a lower degeneracy (two angles are needed to fix a line) 
than the planar orbits (three angles are needed to fix a polygon). 
Moreover, it does not
play a role for the moments of inertia, which are the main concern of this paper.

The shell corrections to the particle number, $N_{sh}$, and to the energy,
$E_{sh}$, are given by the integrals
\beq\label{NshEsh}
N_{sh}(k_F)=\int^{k_F} g_{sh}(k) dk,
~~~E_{sh}(k_F)=\int^{k_F} \frac{\hbar^2}{2m} (k^2-k_F^2)g_{sh}(k) dk.
\eeq
 Carrying out the integrations
\footnote{The expression $E_{sh}$ approximates Strutinsky's
shell correction energy $E(N)-\tilde E(N)$\cite{str68}. 
It is correct in linear order
of the difference $\mu-\tilde \mu$ between the Fermi energies in the systems
with and without shell structure. The integrations are carried out 
in stationary phase approximation, 
which keeps only the leading term in $1/L_\beta$.
 For a comprehensive discussion see \cite{bba97}, p. 227. The latter does
not include the damping factors $D_\beta$. They are easily taken into
account by the equivalence of first averaging $g_{sh}(e)$ with the function
$f(e-e')$ and  then evaluating the integrals (\ref{NshEsh}) with
first evaluating the integrals and then average $N_{sh}(\mu)$ and
$E_{sh}(\mu)$ with the function $f(\mu-\mu')$.} one obtains:
\beq
N_{sh}(k_F)=-\sum_\beta\frac{\hbar}{\tau_\beta}{\cal A}_\beta(k_F)
\cos(L_{\beta}k_F+\nu_{\beta})D\left(\frac{k_FL_\beta}{\gamma R_o}\right),
\label{Nsh}
\eeq
and
\beq
E_{sh}(k_F)=\sum_\beta\left(\frac{\hbar}{\tau_\beta}\right)^2
{\cal A}_\beta(k_F)
\sin(L_{\beta}k_F+\nu_{\beta})D\left(\frac{k_FL_\beta}{\gamma R}\right)
=\sum_\beta\left(\frac{\hbar}{\tau_\beta}\right)^2{g_\beta(k_F)}\label{Esh}.
\eeq
The period of revolution $\tau_\beta$  
of a particle moving  with
the Fermi momentum $\hbar k_F$ on the orbit $\beta$ is given by
\beq\label{T}
\frac{\hbar}{\tau_\beta}=\frac{\hbar^2k_F}{mL_\beta}=
\frac{2e_F}{b}\frac{R_o}{L_\beta}N^{-1/3}.
\eeq
The shell energy is $E_{sh}\sim e_FN^{1/6}$ as compared with 
the smooth part of the Fermi gas, which is $\sim e_F N$.

Eqs. (\ref{Nsh},\ref{Esh}) should be understood in the following way.
Both $N_{sh}$ and $E_{sh}$ depend on the particle number via
$R_o=r_o \tilde  N^{1/3}$, where we have introduced the ``smooth''
particle number for a clean notation. The actual particle number
$N=\tilde N+N_{sh}(\tilde N)$ and the shell energy $E_{sh}(\tilde N)$   
are functions of the parameter $\tilde N$, which then defines $E_{sh}(N)$
in a parametric
form.  However when discussing the extrema of $E_{sh}$, we
need not resort to this sophistication,
 because $\cos(L_{\beta}k_F+\nu_{\beta})=0$ 
where $\sin(L_{\beta}k_F+\nu_{\beta})=\pm 1 $.

\subsection{Basic shell and supershell structure}\label{s:potshell}

Let us discuss the main features of the shell structure of the spherical 
cavity \cite{bbl72} as an educational example. The most important orbits
are the triangle ($\beta=\triangle$) and the square ($\beta=\Box$), to which
we restrict the sums over $\beta$.   
They are the shortest orbits
with the lengths $L_\triangle=5.19R_o$ and $L_\Box=5.66R_o$.
Since, 
$L_\triangle\approx L_\Box$, one has 
$\tau_\triangle\approx \tau_\Box \approx \tau$,  
${\cal A}_\triangle\approx {\cal A}_\Box\approx {\cal A}$
and $D_\triangle\approx D_\Box\approx D$. Using the addition theorem for
the  sine function one finds
\begin{equation}\label{Eshsuper}
E_{SH}=2 \left(\frac{\hbar}{\tau}\right)^2 {\cal A}
\sin{\left(k_FL +\bar \nu\right)}
\cos{\left(k_F\Delta L +\Delta \nu\right)}D,
\end{equation}
with $ L= (L_\triangle+L_\Box)/2=5.42 R$ and
$ \Delta L= (L_\Box-L_\triangle)/2=0.24 R$  and the analogous
definitions for $\bar \nu$ and $\Delta \nu$.
A well known
phenomenon is encountered: the
superposition of two oscillations with similar frequency results in a
beat mode.
The fast oscillation represents  basic shell structure and the slow beat 
mode was called supershell structure \cite{nis90}. 

The phenomenon of supershell structure was observed in Na clusters \cite{bjo90}, almost
three decades after it was predicted \cite{bbl72}. It is realistic 
to assume that the conduction electrons move in a cavity.
The basic shell closures correspond to the minima of the sine function,  
which lie at $k_FL+\bar\nu=\pi(2 n + 3/4)$. With $ L=5.42 R_o$ and $k_Fr_o=b$
 one finds that the closed shells appear at $ N^{1/3}= 0.60 n +c$. 
Overall, the   experimental magic numbers  in Na clusters are  well
reproduced by the relation $ N^{1/3}= 0.61 n +0.50$ \cite{bjo90}.
The magic numbers 58, 92, 138, 198  give $92^{1/3}-58^{1/3}=0.64$,
$138^{1/3}-92^{1/3}=0.65$, and $198^{1/3}-138^{1/3}=0.66$. The 
somewhat longer period for the small clusters as compared to  the POT 
value for the cavity is due to the assumption of a cavity instead of 
a potential of finite depth and surface thickness.

The slow oscillation is the supershell structure and has a half period of
$k_F\Delta L=\pi$, which corresponds to $L/2\Delta L \approx 12$ shells. In
the experiment, the beat minimum appears around $n=15$ \cite{bjo90}.
When the cosine function of the slow oscillation changes sign, the maxima of
the fast oscillation become minima, i.e., the  new
shell closures, which are shifted by half a shell as compared with 
the ones in the
lower beat. This phase shift was also observed at $n=15$ \cite{bjo90}.  
A more careful
application of POT than given here (see \cite{bbl72,nis90}) 
removes the discrepancy between the calculated and observed shell number
 where the beat minimum appears. 

\subsection{Deformation}\label{s:potdef}

In the middle between the closed shells, $E_{sh}>0$ for spherical shape.
 Nuclei and alkali clusters reduce this shell energy by taking 
a non-spherical shape,
i.e., they avoid a high level density near the Fermi level. This is analogous 
to the Jahn-Teller effect in molecules. Due to symmetry, the electronic levels 
of a molecule may be degenerate. 
If such a degenerate level is incompletely filled, 
the molecule changes its shape such that the degeneracy is lifted. 
Whereas the final shape of the molecule is determined by the balance between
this driving force and the restoring force of the chemical bonds, in
the case of nuclei and clusters a shape is attained that minimizes the 
level density near the Fermi surface. The corresponding gaps
in the single-particle spectrum are referred to as `` deformed 
shells'' \cite{bra72}. The optimal shapes are     
described by the few lowest multipoles. This has been known for a long time for 
nuclei, where it is experimentally confirmed.
Also in the case of alkali clusters, several multipoles are needed to describe
the equilibrium shapes (see, for 
example the calculations in \cite{fra96}).
 The calculations of the shapes 
in the present paper include the multipoles necessary for a completely
 relaxed axial shape.
POT permits us to understand some basic features of the deformation. 
However, a full understanding of the interplay between the multipoles 
has not yet been reached (concerning octupoles see \cite{yam00}).      

Our present interpretation
of deformed shapes in terms of POT is based on the analysis of 
the spheroidal cavity. Strutinsky and coworkers
did the pioneering work \cite{str77} by discussing $E_{sh}$ as function
of the particle number $N$ and the ratio $\eta$ 
 of the long and short
semi-axes, which are, respectively:
\beq
R_l=r_oN^{1/3}\eta^{2/3}~~~R_s=r_oN^{1/3}\eta^{-1/3}~~~prolate
\eeq
and
\beq
R_l=r_oN^{1/3}\eta^{1/3}~~~R_s=r_oN^{1/3}\eta^{-2/3}~~~oblate.
\eeq
The volume of the cavity is deformation independent.
Fig. \ref{fig:landscape00} (top) shows the shell energy contours $E_{sh}$ as 
function of the 
deformation parameter $\alpha$ (see \cite{pas71}), 
calculated for the Woods-Saxon potential with a very thin 
diffuseness (a=0.05) and no spin-orbit coupling,
which is practically identical with the cavity. 
In this case, to reach the same accuracy the integration mesh for 
 calculating the Hamiltonian matrix elements was taken $\approx 10$ times 
 finer than for the usual calculation with the parameters from the 
 Tab. \ref{tab:wspar}.
For the considered deformation
range the two deformation parameters are related by  
$\eta\approx\sqrt{(1+|\alpha|)/(1-|\alpha|)}$. 

  The most important
  orbits are the polygons lying in the meridian planes, because they have a
two-fold degeneracy: all polygons in one meridian plane have the same length
and  all meridian planes are equivalent. 
 Again, the triangle and the rhombus are the most important orbits, which 
are shown in Fig. \ref{fig:orbits}.
As for the sphere, their interference generates a beat pattern  
\cite{fri90,mag97}, i.e.,
the shell correction $E_{sh}$ has
 the form (\ref{Eshsuper}), where ${\cal A}$ is different.  
 The basic shell structure is governed by 
$\sin(k_FL_\bot+\bar \nu),$ where $L_\bot=(L_\triangle+L_\diamondsuit)/2$
and $\bot$ indicates the meridian plane. (The reason for the labeling
 becomes clear below in the context of rotation.)
In order to keep the expressions simple, we 
approximate $L_\bot\approx L_\diamondsuit$, following Ref. \cite{str77}.
The length of the rhombi is:
\beq
L_\diamondsuit=4N^{1/3}r_o\sqrt{\eta^{2/3}+\eta^{-4/3}}\left\{
\eta^{1/3}~~prolate \atop
1~~~oblate\right..
\eeq

 The equilibrium shape corresponds to the minimum of $E_{sh}$, i.e. to
\beq
k_FL_\bot +\bar \nu=\pi(2n+\frac{3}{4}).
\eeq    
These lines,  $L_\bot(N,\eta)=const$, are also shown 
in Fig. \ref{fig:landscape00}. 
 As seen, the valleys and ridges follow the 
lines, $L_\bot(N,\eta)=const$. They  are  nearly horizontal on the oblate side,
because the function $\sqrt{\eta^{2/3}+\eta^{-4/3}}$ is roughly constant in the
 interesting range of $\eta$.
However, on the prolate side, constant length corresponds to 
approximately $N\propto \eta^{-1}$, which results in the down-sloping curves.
If one starts from a closed shell and spherical shape ($\eta=1$ or $\alpha=0$)  taking
particles away, it is energetically favorable to follow the valley on the 
prolate side. This is H. Frisk's explanation \cite{fri90} for  
the preponderance of prolate over oblate nuclei. The smooth
increase of the deformation with decreasing $N$ is another experimental
fact (see e.g., \cite{bra72})
 which is explained by the down-sloping of lines, $L(N,\eta)=const$,
on the prolate side. 
If $N$ decreases further, the spherical shape in the
next lower valley eventually becomes energetically favored. Since the 
two valleys are separated by a ridge,  the 
deformation decreases abruptly, when the valley near spherical shape  takes 
a lower energy.  This explains the sudden onset of 
nuclear deformation when the open shell is entered. Both features have
been first pointed out in Ref. \cite{str77}. The traverse of the meridian ridge 
is clearly seen in the shell energies of Na clusters
calculated in
Ref. \cite{fra96} (cf. Fig. 2 therein, minimization with respect to $\alpha$ only).

Let us elaborate on the above analysis of deformation from previous work. 
The $N-\eta$ landscape contains more
structure than the sequence of ridges and valleys generated by the  
meridian orbits.
It shows an interference pattern with a second set of valleys and
ridges that is generated by the orbits that lie in the equator plane and
 its neighborhood. The equator orbits are regular polygons as in the
spherical cavity. In contrast to the meridian orbits,
they are only one-fold degenerated with respect to a rotation
around the center. Therefore the valley-ridge structure is less pronounced than
for the two-fold degenerated meridian orbits. Again, the valleys and ridges
correspond to a constant length $L_\Vert=(L_\triangle+L_\Box)/2$, where 
$\Vert$ indicates the equator plane.
 For simplicity we approximate $L_\Vert\approx  L_\Box$, which on 
 the prolate side is given by 
\beq\label{lequ}
L_\Box=4\sqrt{2}r_oN^{1/3}\eta^{-1/3}.
\eeq
 Fig. \ref{fig:landscape00} also shows the lines, $L_\Vert(N,\eta)=const$.
The  ridges start at the maxima of the spherical cavity and the valleys
(not shown) at the respective minima. This is analogous to the meridian orbits, however
the slope of the lines of constant length is positive, since
$N\propto\eta$.  Both types of orbits
originate from the same family at spherical shape. 
The equator ridges and valleys modulate the bottom of the meridian
valleys. If one starts from a closed shell at
$\eta=1$  taking particles away and moves along the bottom of a meridian
valley one passes a hump that is generated by an equator ridge and
continues into a depression generated by an  equator valley.
The depression is the deformed shell closure. There, the meridian and
equator orbits interfere constructively. The equator hump and
depression are clearly seen in the shell energies of Na clusters
calculated in
Ref. \cite{fra96} (cf. Fig. 2 therein, minimization with respect 
to $\alpha$ only).

\subsection{Superdeformation}\label{s:potsuperdef}

Matsuyanagi and coworkers \cite{ari98}
 studied the relation between the periodic orbits and
the appearance of a pronounced shell structure in a prolate cavity with 
an axes ratio of about 2:1, which corresponds to the nuclear superdeformed
shape. They found that different orbits are responsible for the shell
gaps causing superdeformation. The meridian orbits lose importance with
increasing deformation, because they become longer. 
 The radius of the equator plane shrinks with increasing deformation.
Therefore, the equator orbits become shorter and more important.   
Most prominent are the orbits that close only 
after two turns around the symmetry axis ($w=2$ in Eq. (\ref{lenght})).
The five-point star is the  simplest orbit of this type, which is shown in 
Fig. \ref{fig:orbitsd}. Between  
$\eta=1.5$ and 2, these planar orbits become unstable. They
give birth (bifurcation) to three-dimensional orbits with the following path:
during the first turn the orbit is above the equator plane and during the 
second turn it is below. 
Pictures of these three-dimensional orbits
can be found in Ref. \cite{ari98}. The five-point star and the 
double-traverse triangle determine the superdeformed shell structure.
The butterfly orbit shown
in Fig. \ref{fig:orbitsd} is prominent in the deformation interval 
$1.4<\eta<2$. It bifurcates from the double-traverse
diameter orbit in the equator plane. These orbits become more important
than the shorter orbits with only one winding around the center, because 
the amplitude strongly increases near the bifurcation points. 
In fact, the amplitude goes to infinity if one uses the standard 
stationary phase approximation.
A quantitative
description of the enhancement must be based on a more accurate treatment, 
which Magner et al. worked out for the spheroidal cavity \cite{mag01}.  

The lines of constant length of the five-point star orbit in the equator plane
are also included in Fig. \ref{fig:landscape00}. 
The change from the low-deformation to the high-deformation shell structure
is more clearly seen in figures that extend to larger 
deformation (see Refs. \cite{str77,ari98}). However, some
of the superdeformed shell structure is visible on the right fringe of the figure.
 The fact that double-traverse
orbits are responsible for the superdeformed shell structure, is most
directly reflected by the shorter wavelength of the oscillations of the 
shell energy. The period of the spherical shell structure is
$\Delta N^{1/3}=2\pi/(5.4~ b)=0.60$ (cf. Sect. \ref{s:potshell}). 
The double triangle
has the length $10.4~ r_o  N^{1/3} \eta^{-1/3}$, which gives a period
of $ \Delta N^{1/3}=2\pi\eta^{1/3}/(10.4~ b)=0.40$ for $\eta=2$. 
In accordance with this estimate, the quantum calculations 
in Refs. \cite{str77,ari98} give a  ratio of about 1.5  for  the
periods of the spherical and superdeformed shells.  

A closer inspection of Fig. \ref{fig:landscape00} shows that 
at normal deformation ($\alpha \approx 0.3$) the ridges and
valleys from the meridian  orbits seem to be modulated by both the 
single- (triangle, square) and double- (five-point star, 
double traverse triangle) traverse orbits in the equator plane.
Thus, our discussion  in  section \ref{s:potdef} oversimplified
the interplay between meridian and equator
orbits by disregarding the double-traverse orbits, the inclusion of which
accounts for details of the $N - \eta$ landscape in the upper panel 
of Fig. \ref{fig:landscape00} (e. g. the bump at $N=90$ and $\alpha=0.3$).

\subsection{Application to nuclei}\label{s:potappnuc}

The cavity model differs from nuclei in three important aspects:
i) the nuclear potential does not jump to infinity at the surface; ii) there
is spin-orbit coupling; and iii) there are protons and neutrons, which fill the 
shells differently.  We will discuss each of these briefly.

i)  Refs. \cite{str77,nis90} compared the shell energies
of the cavity with the ones of the Woods-Saxon potential. 
They found that the  qualitative features
agree rather well.  We are using as a cavity-like model the Woods-Saxon 
potential with the same depth, but with very small diffuseness $(a=0.05)$ 
and no spin-orbit coupling.  The results for such a model are close to the 
true cavity model.

ii) Incorporating the spin-orbit coupling  into  
POT (cf. \cite{fri93,bol99,bra00}) is complicated because the 
spin cannot be treated semiclassically.   
Simple interpretations as for the spinless POT have not been found yet.
 However, the 
shell energies calculated from Woods-Saxon potentials with and without
a spin-orbit part show strong similarities if one compares them at
the same fractional filling of the spherical shells, as seen by comparing
the upper and lower panel of Fig. \ref{fig:landscape00}.
In the following we will assume that this also holds
for the rotational energies.

iii) Depending on the nuclide, protons and neutrons may fill a different
fraction of the shell and contribute in a different way to the net 
shell structure. We show the experimental data and the calculations as
 functions of the neutron number $N$, because the neutron shell structure
is more clearly visible. Around the spherical shell closure
at $N=126$,  $Z$ is not too different from $82$, i.e., the proton and neutron
shell contributions are in phase.  They get progressively
out of phase with decreasing $N$. At $N=82$, $Z$ is around 64, 
i.e., about mid shell. For 
$50 \leq N \leq 82$ the protons are out of  phase with the neutrons. 
The neutron shell contributions are stronger than the proton contributions
in the heavy nuclei. Therefore, the neutron shell structure shows up in
the $N$ dependence of the energies and moments of inertia.
 However, it is somewhat modified due to the  proton shell contribution,
which depends on $Z$. 
For our qualitative discussion, we may 
compare the  total shell contributions in nuclei at a given fractional
filling $N$ of the neutron shell with  the ones
calculated for the cavity at the same fractional filling
$N$.  For $N >100$, where the neutron shell contribution
are not too much out of phase with the proton contribution, one expects 
that the $N$ dependence of the shell effects is about the same as in the
cavity that contains only one kind of fermions. For $N<100$ one expects 
a weaker $N$ dependence, because the protons no longer enhance the neutron
shell structure. In fact, the proton shell structure may counteract and
modify the neutron shell structure by changing the deformation.
A detailed discussion of this interplay goes beyond the scope of this paper.  
However, some features of the shell structure that are a consequence of 
this interplay will be discussed below.
               
The magic numbers 50, 82, 126 (and 184 predicted) correspond to 
periods of the spherical shells  $82^{1/3}-50^{1/3}=0.66$,
$126^{1/3}-82^{1/3}=0.67$, $184^{1/3}-126^{1/3}=0.67$,
 which are  the same as in the small clusters and
 somewhat longer than the  period $\Delta N^{1/3}=0.60$ given by POT
for the  cavity (cf. section \ref{s:potshell}).
The closed superdeformed shells appear at
$N=112$ and 88, which corresponds to a period of
$112^{1/3}-88^{1/3}=0.37$. The POT period of the superdeformed shells 
in the cavity  is $\Delta N^{1/3}=0.40$ (cf. \ref{s:potsuperdef}). 
 The ratio of 1.7 between the experimental periods of 
the spherical and superdeformed shells in nuclei is somewhat larger
than the POT prediction of 1.5. The ratio reflects
the different lengths of 
single-traversed orbits in the sphere and the double-traversed orbits
in the equator plane of the spheroid with the axes ratio 2:1.    

The experimental shell energies at zero spin in Fig. \ref{fig:deex0}
show quite clearly the interference between the equator
 and meridian orbits at normal deformation.  Going down from the closed 
 shell at $N=126$ one climbs up the bottom of the valley generated by the 
 meridian orbits, crossing the ridge generated by the equator orbits, and 
 reaches the deformation region generated by the constructive interference of 
 both the meridian and equator valleys (N=98). One then follows the valley 
 generated by the equator orbits.  Along the bottom of this valley one has 
 to go over the ridge generated by the meridian orbits (N = 90) in order 
 to reach the spherical minimum at  $N=82$.
Though less pronounced, the same pattern is seen in the 
shell $50 \leq N \leq 82$. 

The Fermi gas relations are modified for the nucleus because each orbital
state is occupied by four fermions (proton, neutron, spin up, spin down). 
The radius is $R=r_o A^{1/3}$ with $r_o=1.2~fm$. Assuming 
 $N=Z=A/2$ we have $k_Fr_o=b/2^{1/3}$ and $e_F=34~MeV$. 
Hence, for the action one has 
\beq\label{kFRnuc}
k_FR_o=k_Fr_oA^{1/3}=b\left(\frac{A}{2}\right)^{1/3}=bN^{1/3}, 
\eeq
as in the case of a cavity with only one kind of fermions.
Studying the rotation below, we encounter the inverse energy
\beq\label{2mR2nuc}
\frac{2mR^2}{\hbar^2}=\frac{2m\left(r_oA^{1/3}k_F\right)^2}{(\hbar k_F)^2}
=\frac{\left(b(A/2)^{1/3}\right)^2}{e_F}=\frac{\left(bN^{1/3}\right)^2}{e_F}.
\eeq
In the heavy nuclei, where $N>A/2$, there appears the ambiguity, whether 
one should use for $N$ the neutron number or $A/2$. The difference in
the estimates is not significant in the context of our qualitative discussion.
In order to be definite,  we use $N=A/2$ whenever we refer to $A$
and the actual neutron number whenever we refer to $N$.

\subsection{Influence of rotation}\label{s:potrot}

Rotation is taken into account by applying POT to the Routhian
\beq\label{routhian}
H'=H-\omega l,
\eeq
where $H$ is the Hamiltonian of the deformed cavity, $\omega$ the angular
velocity and $l$ the projection of the orbital \am on the axis of rotation. 
The spin is disregarded.
If $\omega$ is interpreted as the Larmor frequency, the  Routhian 
(\ref{routhian}) agrees with the Hamiltonian of a system of electrons in
a weak magnetic field. Hence, we can directly use several results from
studies of a cavity in a magnetic field.

The smooth part of the moment of inertia is the rigid-body value
\footnote{The rigid-body value is only the leading term. The next term
of relative order $N^{-2/3}$ is analogous to the Landau diamagnetism, which
is caused by surface currents. See e.g. \cite{ecaja}.}, which is for
the spherical cavity containing only one kind of fermions
\beq\label{Jrig}
{\cal J}_{rig}=\frac{2}{5}mR^2N=\frac{b^2\hbar^2}{5e_F}N^{5/3}.
\eeq
In the case of nuclei it becomes
\beq\label{Jrig}
{\cal J}_{rig}=\frac{2}{5}mR^2A=\frac{b^2\hbar^2}{ 2^{2/3}5e_F}A^{5/3}.
\eeq

The classical orbits in the cavity are modified by the rotation. The particle
moves on a curved trajectory between the reflections on the surface.
At the Fermi level, the deviation from the straight line is proportional 
to the ratio of the 
velocity of the cavity {\bf $v_r=r\omega$} and the particle velocity 
in the non-rotating cavity $v_F=p_F/m$. Using the maximal value of 
{\bf $v_r$} at the surface, we have
\beq
\frac{v_R}{v_F}=\frac{m R \omega}{p_F}=\frac {m r_o N^{1/3}\omega}{p_F}
=\frac{b \hbar \omega }{ 2e_F}N^{1/3},
\eeq   
where we used the Fermi gas estimates. 
From the maximal angular momentum $I_{max}$ 
we consider in this paper
 (c.f. Fig. \ref{fig:imax})  and the experimental \momis $\J_{exp}$
one has $\hbar \omega < I_{max}/\J_{exp}$, which means that 
the ratio {\bf $v_R/v_F< 0.1$}. Therefore, the Cranking
term $\omega l$ can be treated in a perturbative way.
 Kolomietz et al. 
\cite{kol79} applied the perturbation theory to a rotating spherical cavity.
Tanaka et al. \cite{tan96} and Frauendorf et al. \cite{fra98} used it for 
studies of the magnetic response of electrons in a spherical cavity.    
Creagh \cite{cre96} formulated the perturbative approach in a general way. 
The method and the treatment of a weak magnetic field are exposed in detail 
in \cite{bba97}, which we follow here.

The change of the action due to rotation is given in first order by
\beq\label{phi1}
\Delta S_\beta = \omega \int_\beta l dt = \omega \tau_\beta l_\beta 
\equiv \hbar \Phi_\beta. 
\eeq
The integration runs over the unperturbed orbit in the non rotating cavity.
In the case of the spherical cavity, the \am 
 $l_\beta$ of the orbit is conserved and the 
integration is trivial. In case of the spheroidal cavity, $l_\beta$ is
conserved for the equator orbits but not for the meridian. For the latter,
$l_\beta$ is the average angular momentum of the orbit which must be found
by evaluating the integral (\ref{phi1}). Since 
$\omega l=m (\vec r \times \vec v)\cdot \vec \omega$, 
\beq\label{phi2}
\hbar \Phi(\theta) = m\int_\beta (\vec r \times d\vec s)\cdot \vec \omega=
2m\int_\beta d\vec f \cdot \vec \omega=2m A_\beta \omega \cos \theta
\equiv \hbar \Phi_\beta \cos\theta, 
\eeq
where $A_\beta$ is the area enclosed by the orbit and $\theta$ is the angle
between the normal of its plane and the axis of rotation. 
Hence, $\Phi$ is the ``rotational flux'' in units of $\hbar$,
of the vector field $2m\vec \omega$ enclosed by the orbit. In the analogous
case of particles in a magnetic field, $\Phi_\beta$ is the magnetic 
flux in units
of the elementary flux quantum.  
   
The rotation manifests itself in the appearance of an additional modulation
factor ${\cal M}$ in the expressions for the level density (\ref{gbeta})
and the shell energy (\ref{NshEsh}). The modulation factor is given by
the average of $\exp(i\Phi)$ over all orbits of the same length, which
belong to the  family $\beta$. In the case of a spherical cavity
\cite{kol79,tan96}, the modulation factor becomes 
\beq\label{msph}
{\cal M}_{\bigcirc}(\Phi_{\beta})=j_o(\Phi_{\beta}),
\eeq
 where $\Phi_{\beta}$ is the flux through
the orbit perpendicular to the rotational axis (or the magnetic field) and
$j_o$ the spherical Bessel function. 

In the case of the spheroidal cavity,
we distinguish between the rotational axis being parallel or 
perpendicular to the symmetry axis. For parallel rotation, only the 
equator family carries rotational flux. It consists of all 
polygons rotated around the 
center. Since $\Phi$ is independent of the rotational angle one has 
\beq
{\cal M}_{\Vert}(\Phi_{\beta}) = \cos (\Phi_{\beta_\Vert})= 
\cos (2m A_{\beta}
 \omega/\hbar)=\cos(\omega \tau_{\beta} l_{\beta}/\hbar).      
\eeq
The cosine function appears because in each family 
the orbits appear in time reversed pairs
(the particles run clockwise and counterclockwise on the orbit).  
The case is analogous to spherical quantum dots (two-dimensional circular
potential  pockets that confine electrons on a surface) in a perpendicular
magnetic field. The modulation factor has been observed as oscillations
of the electric current through the dot, which oscillates with the magnetic
 field strength, where the period of the oscillations is given by the area of
the triangle (see \cite{bba97,rei96}).  

For the meridian orbits
\beq
{\cal M}_{\bot}(\Phi_{\beta}) =\frac{1}{2\pi}\int _0^{2\pi}e^{i\Phi_{\beta}
\cos \theta}d\theta=J_o(\Phi_{\beta}),
\eeq
where $J_o$ is the Bessel function and $\Phi_{\beta}$ the flux
through the orbit perpendicular to the axis of rotation. 
Since the area enclosed by the family members   
lying in one plane differs for the meridian orbits one must average 
the modulation factor over all these orbits (see below). 
Fig. \ref{fig:mod}
shows the three types of modulation factors. 

In the analysis of the experiment and the microscopic calculations we assume
that the rotational energy has the form $\omega^2 \J/2$, which corresponds
to an expansion of ${\cal M}$ up to second order in $\Phi_\beta$.
 Using the expansions of $\cos (\Phi_\beta)$,
$J_o(\Phi_\beta)$ and $j_o(\Phi_\beta)$, one finds
\beq\label{mod}
{\cal M}=1-a\Phi_\beta^2,
\eeq   
where,  respectively,
 $a_{\Vert}$=1/2 and $a_{\bot}$=1/4 for the equator and meridian orbits in the 
spheroidal cavity and $a_{\bigcirc}$=1/6 for the spherical cavity.
One may derive this result directly by expanding $\exp(i\Phi_\beta\cos\theta)$
before integration. The leading term is quadratic, because the linear term 
becomes zero when averaging. The next term is $(\Phi_\beta~ cos \theta)^4/24$, 
which can be neglected in our qualitative discussion as long as
$\Phi_\beta <1.5/\left(\overline {(\cos \theta)^4}\right)^{1/4}$. As seen 
in Fig. \ref{fig:mod}, $1-{\cal M}_\beta$  
deviates from its quadratic approximation by less than 35\%
 for $\Phi_\beta<2$, where the quadratic approximation is better for 
$J_o(\Phi_\beta)$ and $j_o(\Phi_\beta)$ than for $\cos(\Phi_\beta)$. 

Inserting the quadratic approximation (\ref{mod}) of the modulation 
factor into eq. (\ref{Esh}) one obtains:
\beq\label{eshquadom}
E'_{sh}(k_F)=E_{sh}(k_F)-\frac{\omega^2}{2}{\cal J}_{sh}(k_F),
\eeq
and
\beq
{\cal J}_{sh}(k_F)=\sum_\beta 2 a l_\beta^2
{\cal A}_\beta(k_F)
\sin(L_{\beta}k_F+\nu_{\beta})D\left(\frac{k_FL_\beta}{\gamma R}\right)
\label{Jsum}.
\eeq
This low-$\omega$ version has been used in by Frauendorf et al. \cite{fra98}
 for analyzing
the magnetic response of spherical metal clusters and circular quantum dots.
In the present paper we have added the meridian orbits. 

The \am of the orbit is
\beq\label{l}
l_\beta=\frac{2m\bar A_\beta}{ \tau_\beta}=\frac{2\hbar k_F \bar A_\beta}{L_\beta}
=\frac{2\hbar\bar A_\beta}{RL_\beta}bN^{1/3},
\eeq
where $\bar  A_\beta=\sqrt{\overline{A^2}}$ and $\overline{A^2}$ denotes 
the average over the degenerate orbits in one and the same plane.
The exact form of this average will not be derived in this paper, because
it is not important for our qualitative discussion, as seen below.
Working it out, would demand a substantial sophistication of POT. 

The area of the regular polygons in one of the planes of the sphere is
\beq\label{area}
\bar A_{v,w}=A_{v,w}=\frac{v}{2}\sin (\frac{2\pi w}{v})R^2=
\frac{v}{2}\sin (\frac{2\pi w}{v})r_o^2N^{2/3}.
\eeq

The area of the polygons in 
the equator plane of the prolate spheroid is given by the
 same expression containing
$R_s$ instead of $R$. Averaging over the different orientations in the
 equator plane does not give anything new because the area is constant.
  For the basic shell structure, we take 
the mean of the contribution
from the triangle and the square,
\beq\label{apar}
A_\Vert=\frac{1}{2}(A_\triangle+A_\Box)=1.64r_o^2N^{2/3}\left\{
\eta^{-2/3}~~~~~prolate \atop
\eta^{2/3}~~~oblate\right.,
\eeq
where $\Vert$ indicates that the rotational axis is parallel to the 
symmetry axis. 
In the case of the meridian orbits, the areas are rather  complicated
expressions in terms of the half-axes (see \cite{kol79}).
Let us consider the rhomboidal orbits. 
The orbit, whose
diagonals coincide with the axes of the ellipse, encloses the area
\beq\label{arhombi}
A_\diamondsuit=2R_lR_s=2r_o^2N^{2/3}\left\{
\eta^{1/3}~~~~~prolate \atop
\eta^{-1/3}~~~oblate\right..
\eeq  
The area  enclosed by the orbit consisting of lines parallel to 
the axes of the ellipse
differs from (\ref{arhombi}) by the factor $2/\sqrt{1+\eta+1/\eta}$. The  
areas of the other rhombi fall in
 between these limits. Since for $\eta<1.5$ the differences
 between the areas are less than
2\%, we use (\ref{arhombi}). 
We assume that the area of the triangles changes with deformation as
given by Eq. (\ref{arhombi}) and use the approximation
\beq\label{aperp}
A_\bot=\frac{1}{2}(A_\triangle+A_\diamondsuit)=1.64r_o^2N^{2/3}\left\{
\eta^{1/3}~~~~~prolate \atop
\eta^{-1/3}~~~oblate\right.,
\eeq  
 where $\bot$
 indicates that the axis of rotation is perpendicular to the symmetry axis. 

 Let us now discuss the accuracy of the
quadratic approximation for the modulation factor in the case of nuclei,
which is illustrated in Fig. \ref{fig:mod}.
 The rotational flux  in the sphere is (cf.
 (\ref{2mR2nuc},\ref{phi2},\ref{apar}))
\beq\label{phi}
\Phi=\frac{2m(1.64R_o^2) \hbar \omega}{\hbar^2}=
\frac{1.03 b^2 A^{2/3} \hbar \omega}{e_F}.
\eeq
In the microscopic calculations, we use $\hbar \omega=0.3 ~MeV$, which
corresponds to fluxes of $\Phi=0.6$ and 1.1 for  $A=80$ and 200,
 respectively. Hence, the quadratic approximation is rather accurate.
In the case of the experiment, we extract the 
shell contributions to the \momi mostly from the last $10\hbar$, i.e.,
the average spin $\bar I\approx I_{max}-5$. Using the experimental moments of inertia,  
we may estimate the frequency as $ \hbar\omega=\bar I/\J_{exp}$, which
 gives $\hbar \omega = 1.0,~0.5,~0.4~MeV$
and $\Phi=2.0,~1.6,~1.5$
for the regions $A=80,~160,~200$, respectively.  
Hence, for the bulk of the 
data the quadratic approximation is not very accurate, but still acceptable
for our first systematic analysis. It becomes problematic for the light nuclei
and for some the cases that reach very high spin (cf. Fig \ref{fig:imax}). 
Including the deformation reduces the flux for rotation about the symmetry
axis. With $\eta=1.3$, one finds  $\Phi_\Vert=1.34$ for $A=160$, for which
the quadratic approximation of $\cos (\Phi)$ becomes quite reasonable.
The deformation increases  the flux for rotation perpendicular 
to the symmetry axis. With $\eta=1.3$, one finds $\Phi_\bot=1.8$ 
for $A=160$. A parabola
that approximates ${\cal M}(\Phi)$ at large values of $\Phi$ has a lower
curvature than the parabola that approximates the low-$\Phi$ part
(see Fig. \ref{fig:mod}).
Hence, the deviations of ${\cal M}(\Phi)$ from the quadratic form will
tend to reduce the shell contributions to the \momis that we derive from the 
data.

\subsection{Moments of inertia}\label{s:potmomi}

The moments of inertia are given by:
\beq
{\cal J}_{sh\Vert}=l_\Vert^2{\cal A}(k_F)
\sin(L_{\Vert}k_F+\bar\nu)D\left(\frac{k_FL_\Vert}{\gamma R}\right),
\label{momipr}
\eeq
and
\beq
{\cal J}_{sh\bot}=\frac{1}{2}l_\bot^2{\cal A}(k_F)
\sin(L_{\bot}k_F+\bar\nu)D\left(\frac{k_FL_\bot}{\gamma R}\right).
\label{momipp}
\eeq 
The ratio (c.f. Eqs. (\ref{amp},\ref{l},\ref{Jrig}))
\beq
\frac{{\cal J}_{sh}}{{\cal J}_{rig}}\sim   l^2 {\cal A} e_FN^{-5/3}
\sim (N^{1/3})^2N^{5/6}N^{-5/3}= N^{-1/6}
\eeq
is of the order of one, i.e. the shell contribution to the \momi is comparable
 with the smooth rigid-body value.

One may rewrite  Eqs. (\ref{momipr},\ref{momipp})  as follows:
\beq
{\cal J}_{sh\Vert}=l_\Vert^2g_{sh\Vert}(k_F)
\label{momiprg}
\eeq
and
\beq
{\cal J}_{sh\perp}=\frac{1}{2}l_\perp^2g_{sh\perp}(k_F).
\label{momippg}
\eeq 
This is a more quantitative statement of the relation between the level
density and the moment of inertia, which we discussed in Sect. \ref{s:qmcd}.
The deviation of the \momi from the rigid-body value
is not determined by the total deviation of the level density from
the Fermi gas value but only by the  part generated by the orbits
carrying rotational flux. This is indicated by the subscripts
$\Vert$ and $\perp$ for the parallel and perpendicular  orientation
of the rotational axis.  
 
  Taking into account Eq. (\ref{Esh}), one may also relate
 (\ref{momipr},\ref{momipp}) to the ground-state shell energy: 
\beq 
{\cal J}_{sh\Vert}=
\frac{\hbar^2}{e_F^2}(k_F^2 A_\Vert)^2 E_{sh\Vert},
\label{momipre}
\eeq
and
\beq
{\cal J}_{sh\perp}=
\frac{\hbar^2}{2e_F^2}(k_F^2A_\perp)^2 E_{sh\perp},
\label{momippe}
\eeq 
which gives $ 
\J_{sh}\sim \left(\frac{\hbar b^2}{2e_F}\right)^2 N^{4/3} E_{sh}
\approx \frac{\hbar^2}{400~MeV^2}N^{4/3} E_{sh}\approx
 \frac{\hbar^2}{1000~MeV^2}A^{4/3} E_{sh}$.
This is why in Figs. 
~\ref{fig:deex0}, ~\ref{fig:deex20}, and ~\ref{fig:deca30}, the 
shell energy was scaled by a factor $A^{4/3}$.  The scaling 
factor  accounts for 
respective scales on the plots of ${\cal J}_{sh}$ 
and $E_{sh}$, 
which is remarkable considering the qualitative nature of 
our arguments. As in the case of the level density, only the part of
the shell energy originating from the orbits enclosing rotational flux
contributes to the shell \momi.    

The terms  in the 
sum (\ref{Jsum}) giving ${\cal J}_{sh}$  contain the additional factor 
$(\tau_\beta l_{\beta})^2$ (as compared to the terms in the sum 
(\ref{Esh}) giving $E_{sh}$) which is proportional to $A_\beta^2$,
the square of  the 
area enclosed by the classical 
orbits (cf. discussion in the preceding paragraph).
This factor tends to favor the contribution of the longer 
orbits to the moment of inertia as compared to the 
shell energy, which will  introduce more fluctuations in 
the variation of the shell moment of inertia as a function of
neutron number. These fluctuations would be damped by averaging
over nuclei (which we have not done) because  the damping
factor in the sums would  suppress the longer orbits. 

The relations (\ref{momipre},\ref{momippe}) allow us to discuss the gross
shell structure of the \momis with the help of Fig. \ref{fig:landscape00}.
The shell structure of the \momi was first discussed in \cite{pas74}.
Fig. 2 of Ref. \cite{pas74} shows the neutron part of $\J_\perp$ for 
$\alpha=0.3$. There is a maximum below $N=90$, a minimum near $N=106$, and 
another maximum at $N=120$, which correspond, respectively, 
to the ridge, valley, and ridge, generated by the meridian orbits in Fig
\ref{fig:landscape00} (bottom). Fig. 3 of the same paper shows the proton part
of $\J_\perp$ for $\alpha=0.3$. There is a maximum at $Z=58$ a minimum
at $Z=68$ and another maximum above $Z=80$, which correlate well
with the sequence ridge, valley, ridge, 
generated by the meridian orbits in Fig. \ref{fig:landscape00} (bottom).     
The interpretation of Fig. \ref{fig:dmomca} and its experimental
counterpart Fig. \ref{fig:dmomex} is more complicated, because the 
deformation and the orientation of the rotational axis are optimized.
For this reason, we first discuss the four subsets of calculations shown
in Fig. \ref{fig:dmomcafix}, in each of which the axis of rotation and the
sign of $\alpha$ are fixed,  with the help of Fig. \ref{fig:landscape00}. 
%and Fig. \ref{fig:landscapeso}, left. 

For prolate deformation and rotation
perpendicular to the symmetry axis the meridian orbits
determine the shell moment of inertia, because only they enclose rotational 
flux.  Starting at the $N=82$ minimum and following the east path of lowest 
elevation around the $N=100$ mountain, one goes over the meridian pass, 
which shows up as the maximum of the $\J_{sh \perp}$, arriving at the deformed 
minimum around $N=98$.  Then one follows the valley to the $N=126$ minimum, 
realizing that there is no $N=106$ ridge, which is generated by the equator 
orbits.  Accordingly, $\J_{sh \perp}$ decreases until the minimum at
$N=126$ is reached. 

For rotation about the symmetry axis,
$\J_{sh \Vert}$ is determined by the equator orbits. Following the same path,
one goes through a valley to a minimum at $N=98$ and over a 
hump at $N=106$
(the ridges generated
by the equator orbits are dashed in Fig. \ref{fig:landscape00}.), 
which determine the minimum and maximum of
$\J_{sh \Vert}$, respectively. The behaviour near
$N=82$ will be explained below. For oblate deformation one goes on the
 west path  around the $N=100$ mountain (Fig. \ref{fig:landscape00}), 
 where $\alpha$ reaches values
between -0.15  and -0.20. The maximum of $\J_{sh \perp}$
at {\bf $N\sim 105$} is caused by the meridian ridge and the maximum of $\J_{sh \Vert}$
at $N=90$ reflects the equator ridge. 
  
All four calculations look similar around $N=82$. There is a minimum at
$N=82$ enclosed by two spikes at $N=80$ and $N=86,~88$. Inside this
interval the shape is spherical. When $N$ goes outside  the interval,
 the deformation
 suddenly jumps to a substantial value and the shell contribution drops. 
The deformation jump is a 
consequence of the topology of the map, i.e., Fig. $\ref{fig:landscape00}$.
Near the closed shell, the equipotential lines are ellipse-like with 
the center at the spherical
minimum. For given $N$, the energy minimum lies at $\alpha=0$.
In the middle of the shell, the equipotential lines become ellipse-like
around the mountain top. Hence, when $N$ increases from 82, $\alpha$ stays
zero as long as the contour lines curve downwards. When the curvature changes
sign, $\alpha$ jumps to a substantial value ($|\alpha|\sim 0.15$).
The analog happens when $N$ decreases  from its magic value, 82.
Hence within the interval $80\leq N \leq 88$, the \momis reflect the parabolic
$N$-dependence
of the shell energy for spherical shape. Outside, the deformed shell structure 
shows up. The distinct structure around $N=82$ is due to the 
interplay between the proton and neutron shell contributions, 
which are out of phase. The isotones  around $N=82$ have 
$60\leq Z \leq 70$, i.e., they have a large positive proton shell energy for 
$\alpha=0$. Accordingly, the protons make a large positive shell contribution
to the moment of inertia, which mostly compensates the negative neutron 
contribution. Therefore, the total shell contribution to the \momi
is only slightly negative at $N=82$ and becomes positive on both
sides. When the deformation jumps to finite values, the 
positive proton shell contribution is drastically reduced, which ends 
the spikes. 

Around $N=126$, the  
protons and neutrons are in phase and the total shell contribution stays
negative. The \momis reflect the shell energy. Going down on the prolate side,
the ridge due to the equator orbits shows up 
in $\J_{sh\Vert}$ as the bump near $N=106$ in Fig. \ref{fig:dmomcafix}, which is not seen in
$\J_{sh\bot}$, because the equatorial orbits enclose no flux.   
Going down on the oblate side, the maximum of $\J_{sh\bot}$
 at $N\approx 100$, reflects the upper side of the mountain in the shell energy,
which is generated by the meridian orbits. The  maximum of $\J_{sh\Vert}$
 at $N\approx 90$ reflects the lower side of the mountain in the shell energy,
which is generated by the equatorial orbits. 

For  $N$ increasing from 
126,  the calculations assuming $\alpha\geq 0$ result in a spherical 
equilibrium shape. The same is true for the calculation constrained to
$\alpha\leq 0 $ and perpendicular orientation of the rotational axis.
In these cases the shell \momi has the 
deep minimum at $N=126$ characteristic for spherical shape. In contrast to 
the $N=82$ region, where protons and neutrons are not in phase, the 
proton number here is $Z\approx 82$, in phase with the neutrons. The combined 
contribution to 
 $\J_{sh}$ leads to the strongly negative values, which reach almost  
the  \rig value ($\J_{calc}$ becomes nearly zero). 
 The calculation
that assumes $\alpha \leq 0$ and parallel orientation of the rotational axis 
gives  $-0.15 < \alpha < -0.20$ for most nuclei. 
The shell \momis are  much less negative
 than for a spherical shape, as the comparison with the other 
calculations in Fig. \ref{fig:dmomcafix} shows. Only for $Z=82$ 
and $N\leq 120$ is the shape spherical,
and then the shell \momi decreases strongly. We discuss the competition
 between the shapes  and different orientations of the rotational 
axes in section \ref{s:potsefs}.    

Fig. \ref{fig:dmomca}  is a combination of Figs. \ref{fig:dmomcafix}.
Near $N=82$, it shows the two spikes at $N=78$ and $N=86$, 
which are caused by the sudden transition from spherical to deformed shape.
For $90\leq N < 100$ the shape is prolate and the rotational axis
 perpendicular to the symmetry axis. The shell \momi
 $\J_{sh\bot}$ gradually 
decreases from about 20 $\hbar^2/MeV$ to -10 $\hbar^2/MeV$. Above 
$N=102$, rotation about the symmetry axis is preferred, and the peak
of $\J_{sh\Vert}$ at $N=106$ shows up. Above
$N=110$, the shape is oblate and the rotational axis
parallel to the symmetry axis. Only the $Z=82$ chain has a spherical
shape at $\hbar \omega=0.3~MeV$. 

The experimental shell \momis have a similar $N$ dependence as the 
calculated ones. The $N=106$ peak is less pronounced than in the calculations.
This can be understood as follows. In the calculation it is assumed that
the rotational axis is parallel to the symmetry axis, i.e., the yrast line
is a sequence of high-K band heads. The experimental yrast lines contain only
few such states. Most of the states entering the fit are band members 
with $I>K$. For such states, the rotational axis is neither parallel nor
perpendicular to the symmetry axis, rather it has an
 intermediate orientation. As a consequence, the shell \momi should 
be between  $\J_{sh\Vert}$ and $\J_{sh\bot}$, i.e. 
smaller than $\J_{sh\Vert}$. In our analysis of the experiment, we
subtract the \rig \momi for rotation perpendicular to the symmetry axis.
For a tilted rotational axis the \rig value will be smaller and $\J_{sh}$
larger. 

\subsection{Shell energies at finite spin}\label{s:potsefs}

Fig. \ref{fig:landscape0306} shows the shell energies of a
 cavity rotating perpendicular to its symmetry axis. Lets us first consider
spherical shape ($\alpha=0$). The modulation factor 
${\cal M}_{\bigcirc}(\Phi)$ given by eq. (\ref{msph}) damps the 
shell structure of the non-rotating cavity. The flux through the 
orbits that govern the basic shell structure (mean of the triangle and square)
  is
\beq\label{flux}
\Phi=1.64 b^2N^{2/3}\hbar \omega/e_F=0.178~(MeV)^{-1}N^{2/3}\hbar \omega,
\eeq
 which gives 1.26 for $N=115$ and $\hbar \omega=0.3~MeV$. As seen in Fig. 
\ref{fig:mod}, the modulation factor 
${\cal M}_{\bigcirc}(1.25)=0.76$. The top of the 
$N=115$ mountain is at 
$18~MeV$ for $\omega=0$ (cf. Fig. \ref{fig:landscape00}) and $11~MeV$
for $ \hbar \omega=0.3~MeV$, which corresponds to a reduction of 0.6. For
$ \hbar \omega=0.6~MeV$ one has $\Phi=2.5$ which gives 
 ${\cal M}_{\bigcirc}(2.5)=0.24$. The above maximum energy is $2~MeV$ for
this frequency, corresponding to a reduction factor of 0.1. Similar
qualitative agreement holds for the other minima and maxima at $\alpha=0$.
In the case of the realistic potential (cf. Figs. \ref{fig:landscape00} and
\ref{fig:landscapeso}), the reduction of the shell energy
in the quantal calculation is 0.81, 0.70, and 0.73 for 
$\hbar \omega=0.3~MeV$ and $N$=82, 100, and 126,
 which is to be  compared with the
POT estimates 0.84, 0.79, and 0.73,  respectively.

Comparing the experimental shell
energy at $I=20$ (Fig. \ref{fig:deex20}) with the ground-state shell
energy (Fig. \ref{fig:deex0}), one sees that the depth of the minimum at neutron
number $N=126$ is reduced by about a factor of 2.
A flux of $\Phi=1.9$ is needed to make the modulation factor equal to 0.5.
The estimate of
the rotational frequency is not very accurate, because of the irregular
sequence of the yrast levels.  
Near $N=126$, the experimental \momis scatter
between 30 and 45 $\hbar^2/MeV$, 
which gives $\hbar \omega =0.66 - 0.44~MeV$ and $\Phi=2.95 - 1.96$
for $I=20$.   
Another possibility to estimate the frequency is to average the level
spacings $(E(I)-E(I-2))/2$ over several spins near $I=20$. 
The sample of nuclides, for which enough data exist, is $(Z,N)$=(82,120),
(82,122),  (86,123), (86,126),  (86,128), (88,126), (88,128), (88,129), and
(88,130). 
The distribution of the frequencies is $\hbar\omega=0.36\pm 0.22~MeV$,
which corresponds to the flux {\bf $\Phi=1.6\pm 0.9$}.
Hence, the estimated frequencies and reduction of the shell structure
are consistent. However, more data are needed to improve the 
statistical significance.

Let us now consider  a substantial deformation. Only the orbits
 that carry flux enter the modulation factor, and their contribution to
the shell energy changes with $\omega$. The contribution of the other 
orbits remains the same. This simple observation is the key for understanding 
the evolution of shell energies with frequency.
Figs. \ref{fig:landscape0306} and \ref{fig:landscapeso} (lower panels)
show the shell energy for rotation perpendicular to the symmetry axis.
With increasing frequency $\omega$, the valley-ridge system generated by
the meridian orbits is attenuated and mostly gone at $\hbar \omega=0.6~MeV$.
What remains is the upsloping valley-ridge system generated by the equator
orbits. For $N=100$, $\alpha=0.3$,
 and $\hbar \omega=0.6~MeV$, the flux $\Phi_\bot=2.5$ 
(cf. Eqs. (\ref{flux},\ref{aperp}))
which gives  ${\cal M}_\bot(2.5)=-0.06$ (cf. Fig. \ref{fig:mod}). 
The contribution of the
meridian orbits has essentially vanished. For rotation parallel to the symmetry
axis, only the  contribution from the equator orbits is modified.
For $N=100$, $\alpha=0.3$,
and $\hbar \omega=0.6~MeV$, the flux $\Phi_\Vert=1.9$ 
 (cf. Eqs. (\ref{flux},\ref{apar}))
which gives  ${\cal M}_\Vert(1.9)=-0.3$ (cf. Fig. \ref{fig:mod}).
Now the equator orbits contribute with the opposite sign, i. e. valleys
become ridges and vice versa. The $N=100$ mountain, which for
$\omega=0$ is generated by the constructive interference of the 
meridian ridge and the equator ridge, becomes a saddle due to the destructive 
interference. For the same reason, the $N=126$ minimum also becomes 
a saddle. The topology does not change for $N=50$ because the fluxes 
are smaller.

In principle, the argument given in the preceding paragraph 
is not applicable to 
the  region of small $\alpha$, because it is based on the assumption of
the two isolated  families of equator and meridian orbits.        
Nevertheless, it describes the behavior in this region properly. 
The reason will be
given in the appendix, where we shall discuss the region of 
small deformation.     
It will be shown that the main features of Figs. \ref{fig:landscapeso}
can be understood in terms of one family of tetragonal orbits, if the 
integral over the different orientations of these orbits is exactly
evaluated. 

Without rotation, the minima and maxima of the shell energy lie on 
the $\alpha=0$ axis, because the spherical symmetry correspond 
to maximal degeneracy of the orbits which is reflected by a maximal 
amplitude of the oscillating terms. Figs. \ref{fig:landscape0306}
and \ref{fig:landscapeso} show that for finite 
rotational frequency the minima and maxima are shifted to slightly negative 
values of $\alpha$ and somewhat smaller values of $N$. We could
 not find
an explanation for this shift within our simple version of POT.

In the deformed nucleus the shell effects depend on the orientation 
of the rotational axis with respect to the deformed potential. We have
studied the most important possibilities that the rotational axis is 
parallel or perpendicular to the symmetry axis. As discussed above,
 the shell contribution
to the rotational energy depends on the flux through the orbit. The 
\momis for rotation about the symmetry axis are determined by the equator 
orbits and the ones for rotation about the an axis perpendicular 
to the symmetry axis by the meridian orbits. We indicated this 
in the expressions (\ref{momipr},\ref{momipp}), which we refer to as the
parallel and perpendicular moment of inertia. 
The nucleus rotates about the axis with the larger moment of inertia
because the rotational energy is smaller.  
Rotation about the symmetry axis corresponds to the generation of \am
by sequential particle-hole excitations. The spacings between the levels 
are irregular, and many isomers appear in the yrast line. If the 
rotational axis is perpendicular to the symmetry axis, the yrast
 levels organize into regularly spaced rotational bands.  
Hence, rotation  selects the contributions of one of two types of orbits, 
which are combined
in the ground state shell energy.

The correlation between the ground state shell energy and 
the shell contribution to the \momi cause a competition between
the rotational and the ground-state shell energies, which tend to have 
the opposite sign. Let us discuss two examples of this competition.

The equator orbits cause the bump (positive shell energy) around 
$N=106$ and $\alpha=0.2$ seen in the meridian valley (negative shell 
energy) of Fig. \ref{fig:landscape00}.  This would correspond to a 
positive $\J_{sh\Vert}$ and a negative $\J_{sh\bot}$.  In this case, the 
difference is large enough to make $\J_\Vert$ larger than $\J_\bot$ so that 
the rotation about the symmetry axis is preferred.  
In fact, the $N=106$ region is known to be rich in $K$-isomers,
which are states with the \am parallel to the symmetry axis. Experimentally,
the parallel rotation is not quite as favored as in the calculations.
The two \momis are about the same, and the yrast line contains 
high-$K$ bands, for which the rotational axis has an
 intermediate orientation.  
 
The other example is the region above $N=126$. Fig. \ref{fig:landscapeso}
demonstrates that oblate shape and parallel rotation is energetically
favorable as compared to prolate shape and perpendicular rotation
(see the right panels, which exaggerate).  The reason is as follows.  
Fig. \ref{fig:landscape00} shows that positive contributions to $\J_{sh}$ 
come from the parallel rotation on the oblate side and perpendicular rotation 
on the prolate side.  The $\J_{sh\Vert}$ is larger because the 
factor $a_\Vert=1/2$ is larger than $a_\bot=1/4$ in expression
(\ref{eshquadom}) or, equivalently, the $\Phi$-dependence of 
$\cal M_\Vert$ is stronger than for ${\cal M}_\bot$ (cf. Fig. \ref{fig:mod}).
The macroscopic moments of inertia reinforce this choice.  
In fact, this region is well known to for high spin isomers, which have an
 oblate shape and the \am aligned with the symmetry axis. Another such
region lies above $N=82$.

\subsection{Rotation of superdeformed nuclei}

The  orbits that cause shell structure of (prolate) superdeformed nuclei 
(cf. Sec. \ref{s:potsuperdef}) do not carry rotational flux
if the rotational axis is perpendicular to the symmetry axis.
This is evident for the orbits in the equator plane. For the 
butterfly orbit in Fig. \ref{fig:orbitsd} one must take into account that
the rotational flux has a sign. If the rotational axis points out of 
the page, it is positive if the particle runs counterclockwise around 
the enclosed area and it is negative if it runs clockwise.  
Therefore, the flux has opposite sign for the two wings of the butterfly
and the total flux is zero. The analogous compensation takes place for the 
three-dimensional orbits. The projections of the 3D five-point star
and of the 3D double-traversed triangle on a meridian plane consist
of two equal areas contributing with opposite 
sign (cf. Fig. 7 of \cite{ari98}) cancelling each other.
In the observed superdeformed rotational bands the axis of rotation is
perpendicular to the symmetry axis. Therefore, their \momis should
be equal to the rigid body value, which is the result of
numerous mean-field calculations (see e.g. \cite{rag93,naz89}).
Fig. \ref{fig:dmomexsd} shows that the experimental deviation from the 
rigid-body value is about 5\% or less in the $A=150$ region. A similar 
analysis of the {\bf $A=190$} region is not possible, because the \momis are not
constant, which likely means the \am where the  pairing is small  has not
 yet been reached. Other regions of "superdeformed" nuclei have not 
 been included in that analysis because the deformation is smaller and it 
 is not clear that the equatorial orbits dominate.  Exceptions 
 are for the nuclei $^{91}Tc$ and $^{108}Cd$ which have an axis ratio close to 
 2 and indeed have a rigid-body moment of inertia.

If the axis of rotation is parallel to the symmetry axis the 
equator orbits carry rotational flux. Since they cause the
 strongly negative shell energy of the superdeformed nuclei, their shell
contribution to the \momi will also be strongly negative. 
Therefore, small \momis for rotation about the symmetry axis are
expected. This means that the appearance of high-$K$ isomers and bands
 is strongly disfavored in superdeformed nuclei. So far no evidence for
this type of rotation has been reported.
        
Concerning the zero-spin shell structure,
 spherical and superdeformed nuclei are similar. 
In both cases the periodic orbits generate a strongly negative 
contribution to the shell energy, which generates the shell gap.
However, their rotational response is different. In spherical nuclei
no direction is preferred. There is only one \momi, which is
roughly proportional to the level density near the Fermi
surface and, hence, small. 
The symmetry axis and the short axes play a different
role in superdeformed nuclei. The \am has  always the direction
of the short axes, which have the largest \momi.
For this orientation, the orbits that cause the shell
energy do not carry rotational flux. Therefore, the \momi takes the 
rigid-body value and the shell structure is not damped by rotation as in 
the case of the spherical closed shells. This example demonstrates that
the simple relation between the level density and the 
\momi holds only for not too large deformation. 
 In superdeformed nuclei, the \momi takes the rigid-body value, 
although there is a substantial shell gap.

\subsection{Finite temperature and unresolved spectra}

The study of the unresolved $\gamma$-continua permits us to explore
regions of finite temperature and large elongation. Since the information
on the rotational response is less direct than in the resolved spectra, one
has to make assumptions about the \momis in  the evaluation of the data.
Therefore, one should know to what extent the \momi is expected to
deviate from the rigid-body value.

Finite temperature causes additional damping of the shell structure, 
which is taken into account by multiplying the zero-temperature 
damping factor $D$ with the temperature damping factor
\beq
D_T=\frac{\pi \tau_\beta T/\hbar}{\sinh (\pi \tau_\beta T/\hbar)},
\eeq
where $T$ is the temperature \cite{kol79,fra98} \footnote{
 Ref. \cite{fra98} contains a misprint. The factor $\pi$ in the 
argument is missing.}.  Its argument is $\pi  \tau T/ \hbar=
\pi m LT/(\hbar^2 k_F)=5.4 \times \pi bTN^{1/3}/(2 e_F)
= 0.38~[MeV]^{-1}A^{1/3}T$ for the orbit length $L=5.4 R$, which 
determines the basic shell oscillations. 
The $\gamma$-cascades of high-spin experiments on nuclei 
in the mass region 170  are characterized by temperatures of 
0.5 $MeV$ or less, which corresponds to \mbox{ $0.84<D_T<1$},
i. e. the \momis are not very different from their zero-temperature values.
The thermal damping becomes important for $T=1~MeV$, where $D_T=0.52$, and the
shell structure is wiped out at $T=3~MeV$, where $D_T=0.02$. These estimates
agree well with the microscopic  calculations of the shell contribution
to the \momis at finite temperature in Ref. \cite{pas74}.   

In the experiments \cite{ste02} rotational frequencies between
0.4 and 0.8 $MeV$ are reached. For moderate
deformation ($\eta=1.3$), the meridian orbits carry rotational fluxes of
1.35 and 2.7, which give modulation factors of 0.6 and -0.1, respectively
(cf. Fig \ref{fig:mod}).
Hence, within the frequency interval the shell structure is expected 
to go to zero and return weakly inverted.
With increasing \am the nuclei reach the transition
 to very elongated shapes. Then the class of 
orbits discussed for superdeformation takes over and the \momis take the
rigid-body value.
 The experiments on the unresolved spectra  are consistent with
weak deviations of the \momis from the rigid-body value \cite{ste02}. 

\subsection{Currents in the rotating frame}

The deviations of the \momis from the \rig value
 indicate that there must be net
currents in the body-fixed frame. By the correspondence principle, one 
expects that these currents are generated by the nucleons on the classical
periodic orbits.  We have not studied the currents 
in any detail, leaving this interesting question for 
later.   In this paper, we give only a heuristic argument.  

For simplicity, let us consider the equator orbits and rotation parallel
to the symmetry axis. 
Classically, the particles revolve the symmetry
axis, where we consider only square orbits for simplicity (to be brief
we omit the subscript $\Box$).
There are two orbits: the particle runs counterclockwise or clockwise,
which correspond to $l>0$ and $l<0$, respectively.
The shell contribution to level density associated with the square
orbits is $g$, therefore each of the two contributes $g/2$.
Without rotation, there is the same number of particles on
both types of orbits, and the net current is zero.
When we rotate the frame, a particle with 
angular momentum $l$ gains  the  energy $-l\omega$. 
Particles will redistribute from the unfavored ($l<0$) to the 
favored ($l>0)$ orbits, and the net \am and current will no longer be zero.
For a given chemical potential $\lambda$, the energy takes a minimum if all 
particles within the energy interval $l\omega$ above $\lambda$
flip from $-l$ to $l$.  
Hence, the number of flipped particles is
$l\omega g/2$.  Since each gains  
$2 l$ of angular momentum,  
the total angular momentum gain is 
$l^2  \omega g$. In agreement with Eq. (\ref{momiprg}), this corresponds 
to the shell moment of inertia of ${\cal J}_{sh} = gl^2$.

This shows that the mass current, and therefore the shell contribution 
to the moment of inertia, comes from nucleons circulating 
on the contributing classical periodic orbits with the Fermi velocity.  
Let us consider the current distribution.
Since the classical mass current of a particle
on the square orbit is 
$m/\tau=p_F/L$, the total mass current due to all
particles on the square orbit is
\beq
J_{sh} = \frac{m}{\tau} l\omega g_{sh}=\frac{p_F}{L} l\omega g_{sh}.
\eeq  
This current is distributed between the outer circle and the inner 
circle tangent to the sides of the orbits in Fig. 
\ref{fig:orbits}.  These circles limit the region accessible for a 
classical orbit.
An exponentially decreasing tail will reach into the classically 
forbidden region. 

The current density is $j_{sh}=J_{sh}/\sigma$, 
where $\sigma$ is the cross section
of the area in which $j_{sh}$ is strong. Perpendicular to the
symmetry axis the classically accessible area extends between 
the surface ($r=R_s$) and $r=R_s/\sqrt{2}\approx 0.7R_s$, i.e., about
0.3 $R_s$. For an estimate of extension parallel to the symmetry axis
a more profound analysis in the framework of POT is needed.
As an order of magnitude estimate we take 0.3 $R_l$.
Using the expression (\ref{amp}) for ${\cal A}_\Box$ for the total mass current 
and considering maximal 
shell contribution (sine function equal to minus one) the current density
$j_{sh}\sim 3.74 m r_o^{-2} A^{1/6} \omega$ for $\eta = 1.3$.  
 The current density for the rigid 
flow is:
\beq
j_{rig}= m\rho r\omega \sim 0.8 m R_s \omega \frac{3}{4\pi r_o^{3}} 
\sim 0.175 m r_o^{-2} A^{1/3} \omega,
\eeq
where we used $r = 0.8 R_s$, which is a rough estimate of the average 
value of r of the particle on the classical orbit.  
The large ratio $j_{sh}/j_{rig}\sim 20 N^{-1/6}$ indicates strong
 surface currents, which are needed to contribute to the total \am
an amount comparable to the contribution from the \rig flow, 
which is distributed over the whole nucleus.  The same type of currents 
will also be generated by
the other polygon-type orbits, both in the equator and meridian planes.  
The shell currents circulate opposite to the rotation if
$g_{sh}<0$, i. e. they reduce the angular momentum. For $g_{sh}>0$ they circulate
with the rotation and increase the angular momentum.

\section{Conclusion}

We have shown that the nuclear moments of inertia at high spins 
along the yrast line differ substantially from the  rigid-body value. 
The differences cannot be attributed to pair correlations, rather they 
manifest the shell structure. Comparing experimental data 
with quantum mechanical mean-field calculations {\it assuming zero pairing}
we find  a similar dependence on the neutron number, which is strongly
 correlated with the well known shell energy at zero angular momentum.
The data  and the quantal calculations can be interpreted using the 
semiclassical Periodic Orbit Theory, which relates the quantal shell 
effects to the characteristics of classical periodic orbits in the 
same potential. A number of features, such as the small \momis near closed shells,
\momis that {\it exceed} the rigid-body value around neutron number 90,
the appearance of isomers near neutron number 106, the correlation between
the deviations of the \momis from the rigid-body values and
 the ground-state shell energies, and the damping of 
shell structure with increasing \am  are explained from this 
new perspective. The gross shell effects persist along the yrast line up to the highest 
observed angular momenta.  The deviation of the mass current
 from the rigid-body flow pattern is
generated by nucleons on classical periodic orbits near the nuclear surface. 
 The Periodic Orbit Theory provides a qualitative description of these 
shell effects in terms of classical mass currents in the rotating 
frame. 

\begin{acknowledgments}
We would like to thank P. Fallon for providing the data base and analysis 
program for superdeformed nuclei, R. Firestone for help in constructing 
figures, and F. Stephens for critical reading of the manuscript.  
This work was supported by the Director, Office of Energy Research,
Division of Nuclear Physics of the Office of High Energy and Nuclear
Physics of the U.S. Department of Energy under contract
No. DE-AC03-76SF00098 and DE-FG02-95ER40934, and by a EU grant, 
INTAS-93-151-EXT.
\end{acknowledgments}

\appendix*

\section{}
  
Here we demonstrate that the salient features of the shell
structure for moderate deformation derive from a simple integral
over the family of tetragonal orbits. We use the perturbative 
approach by Creagh \cite{cre96}. Taking into account the 
change of the length of the orbit in linear order of the 
deformation parameter $\alpha$ (see \cite{mei97})
\footnote{ We ignore the slight difference
between $\alpha$ and the deformation parameter $\epsilon$ 
for our qualitative discussion.}, the contributions of deformation and rotation
to the action are given by 
\bea
\Delta S_{\Vert}/\hbar= -\frac{1}{2} k L_\Box P_2(\cos{\theta})\alpha
-\Phi_\Box \cos{\theta}\\
\Delta S_{\bot}/\hbar= -\frac{1}{2} k L_\Box P_2(\cos{\theta})\alpha
-\Phi_\Box \sin{\theta}\cos{\phi},
\eea
where the Euler angles $\psi,\theta,\phi$ describe the orientation of the 
tetragon. As usual, we denote
 the length of the square in the sphere by $L_\Box$ and the
flux through it by $\Phi_\Box$.  
The changes of the action due to deformation and rotation
 give rise to the modulation factor
\beq\label{Modpert}
{\cal M} =\frac{1}{4\pi}\int _0^{2\pi}\int_0^\pi 
e^{i\Delta S(\theta,\phi)/\hbar}\sin{\theta}d\theta d\phi,
\eeq
which is numerically evaluated. 

Fig. \ref{fig:modna} shows the  shell
contribution to the level density calculated in this way. The similarity with
Figs. \ref{fig:landscape00}, \ref{fig:landscape0306}, and
\ref{fig:landscapeso} is obvious. At large deformation, one may evaluate the
integral (\ref{Modpert}) using the stationary phase approximation.
The derivative of the Legendre polynomial $P_2(\cos{\theta})$ is zero for
$\theta=\pi/2$, which corresponds to the meridian orbits, and for 
$\theta=0,~\pi$, which corresponds to the equator orbits. The structure of
the level density is the consequence of the interference of 
these two families, as discussed in the main text. However, the
interference pattern is still recognizable
at moderate deformations, where the stationary phase approximation becomes
problematic. This justifies our interpretation of the shell structure in
terms of the interference of the meridian and equator orbits.

\pagebreak
\begin{figure}[t]
\vspace{4cm}
%\mbox{\psfig{file=fig1.epsi,width=86mm}}
\caption{Yrast lines for a rotational nucleus $^{160}Er$ (top) and 
	for a non-rotational nucleus $^{150}Dy$ (bottom) together 
	with their fit for the ten highest spins.}
\label{fig:yrast}
\end{figure}
\suppressfloats
\newpage
\begin{figure}[t]
%\mbox{\psfig{file=fig2.eps,angle=270,width=14cm}}
\caption{Yrast lines for a selection of nuclei relative to
their fit for the ten highest spins: $^{160}$Er (diamonds), $^{168}$Hf 
(squares), $^{170}$Hf (triangles) and $^{214}$Ra (crosses).}
\label{fig:pairen}
\end{figure}
\newpage
\begin{figure}
%\mbox{\psfig{file=fig3.ps,angle=270,width=14cm}}
\caption{Experimental moments of inertia 
	 as a function of neutron number for small and normal deformation.  
	 The symbols shown at right indicate the proton number and are the 
	 same for all following figures except figure 5.}
\label{fig:momex}
\end{figure}
\newpage
\begin{figure}[t]
%\mbox{\psfig{file=fig4.ps,angle=270,width=14cm}}
\caption{Experimental deviations from the rigid-body moments of inertia 
	 as a function of neutron number for small and normal deformation.}
\label{fig:dmomex}
\end{figure}
\newpage
\begin{figure}[t]
\vspace{3cm}
%\mbox{\psfig{file=fig5.eps,angle=270,width=14cm}}
\caption{Experimental deviations from the rigid-body moments of inertia 
	 as a function of neutron number for superdeformation.}
\label{fig:dmomexsd}
\end{figure}
\newpage
\begin{figure}[t]
\pagebreak
%\mbox{\psfig{file=fig6.epsi,angle=270,width=14cm}}
\caption{Experimental shell energies at spin 0 as a function of 
	 neutron number.}
\label{fig:deex0}
\end{figure}
\newpage
\begin{figure}[t]
%\mbox{\psfig{file=fig7.ps,angle=270,width=14cm}}
\caption{Experimental shell energies at spin 20 as a function of 
	 neutron number.}
\label{fig:deex20}
\end{figure}
\newpage
\begin{figure}[t]
%\mbox{\psfig{file=fig8.ps,angle=270,width=14cm}}
\caption{Maximum angular momentum observed in each nuclei as a 
	function of neutron number.}
\label{fig:imax}
\end{figure}
\newpage
\begin{figure}[t]
%\mbox{\psfig{file=fig9.ps,angle=270,width=14cm}}
\caption{Same as Fig. \ref{fig:dmomex}, but for the subset of nuclei with maximum 
	  spin greater than $20\hbar$.}
\label{fig:dmomex20}
\end{figure}
\newpage
\begin{figure}[t]
\begin{minipage}[b]{8cm}
\vspace*{-3cm}
\end{minipage}
\begin{minipage}[b]{8cm}
\vspace*{-3cm}
\end{minipage}
\caption{Calculated deviations of the \momi from the rigid-body 
value as a function of neutron number for
different shapes (oblate - left panels, prolate - right panels) 
and different orientations of 
 the rotational axis with respect to the symmetry axis
 (perpendicular - lower panels  and parallel -
upper panels).}  
\label{fig:dmomcafix}
\end{figure}
\newpage
\begin{figure}[t]
%\mbox{\psfig{file=fig11.ps,angle=270,width=14cm}}
\caption{Calculated deviations of the \momi from the rigid-body 
value as a function of neutron number for optimal orientation
of the rotational axis.}
\label{fig:dmomca}
\end{figure}
\newpage
\begin{figure}[t]
%\mbox{\psfig{file=fig12.ps,angle=270,width=14cm}}
\caption{Calculated shell energies at spin 30 as a function of 
	 neutron number.}
\label{fig:deca30}
\end{figure}
\newpage
\begin{figure}[t]
\vspace*{2cm}
\caption{
Classical orbits in the equator plane (upper panel) and the
meridian plane of a normally deformed spheroidal cavity.}  
\label{fig:orbits}
\end{figure}
\newpage
\begin{figure}[t]
\vspace*{1cm}
%\psfig{file=fig14.eps,width=7cm}
%\vspace*{1cm}
\caption{Classical orbits in the equator (upper panel) and the
meridian plane of a super deformed spheroidal cavity.}  
\label{fig:orbitsd}
\end{figure}
\newpage
\begin{figure}[t]
%\vspace*{-3cm}
%\psfig{file=fig15a.eps,width=8cm}
%\psfig{file=T021W00L.EPS,width=12cm}
%\psfig{file=fig15b.eps,width=8cm}
\caption{Shell energy at zero rotational frequency
for two Woods-Saxon potentials.
Upper panel: Cavity-like with small diffuseness $a=0.05 fm$, no spin-orbit potential.  
Lower panel: Realistic parameters from Tab. \ref{tab:wspar} and $(N,Z)=(104,78)$, with 
spin-orbit potential.  It is assumed that for each $N$ corresponds $Z=int(78/104)$.  
The lines of 
constant action for the cavity are included. 
The downsloping lines show the rhombi in the
meridian plane. The upsloping lines starting at
$\alpha=-0.2$ show the squares in the equator plane and the upsloping lines
starting at $\alpha=0.2$ show the five-point star in the equator plane.
The dashed lines correspond to maxima and the full lines to
 minima of the respective shell energies.
The values for the action are chosen such that for 
the equator and meridian orbits the lines go through the minima or maxima
of the shell energy at $\alpha=0$. In the upper panel, 
the action for the five-point star orbit
is chosen such that $L_\star k_F+\nu_\star = \pi(2n+1/2)$, $n$ integer, i.e.,
such that the $\sin$-function in (\ref{gbeta}) is equal  to one. In the 
lower panel a constant is added, which is chosen such that the magic number
$N=88$ for superdeformed shape falls half-way between two lines at 
$\alpha=0.6$.}  
\label{fig:landscape00}
\end{figure}
\newpage
\begin{figure}[t]
\caption{Modulation factors as functions of the rotational flux $\Phi$.
 The dashed lines show the quadratic approximation.}  
\label{fig:mod}
\end{figure}
\newpage
\pagebreak
\begin{figure}[t]
\vspace{8cm}
\caption{Shell energy at finite rotational frequency
for a  Cavity-like potential rotating perpendicular to the symmetry axis. 
Upper panel: $\hbar \omega=0.3~MeV$,
Lower panel: $\hbar \omega=0.6~MeV$.
The lines of constant action are the same as in 
Fig. \protect \ref{fig:landscape00} upper panel.}  
\label{fig:landscape0306}
\end{figure}
%\clearpage
\newpage
\begin{figure}[t]
\begin{minipage}[b]{8cm}
\end{minipage}
\begin{minipage}[b]{8cm}
\end{minipage}
\caption{Shell energy at finite rotational frequency
for a  realistic Woods-Saxon potential with spin-orbit coupling. 
Left panels: $\hbar \omega=0.3~MeV$,
Right panels: $\hbar \omega=0.6~MeV$, Upper panels: rotation 
parallel to the symmetry axis, Lower panels: rotation perpendicular
to the symmetry axis.
The lines of 
constant action  are the same as in Fig. \protect \ref{fig:landscape00} 
lower panel.} 
\label{fig:landscapeso}
\end{figure}
\newpage
\begin{figure}[t]
%\setlength{\unitlength}{1mm}
% \begin{picture}(0,100)
%  \put(-140,160)
%  {
%    { \special
%      {
%        psfile=fig19.eps
%        hoffset=-0
%        voffset=0
%        hscale=100
%        vscale=100
%        angle=-90
%      }
%    }
%  }
%  \end{picture}
\caption{POT level density (arbitrary units) 
 generated by a the family tetragonal orbits in a  
rotating spheriodal cavity. 
Only terms of first order 
in $\alpha$ are taken into account in orbit length.}
\label{fig:modna}
\end{figure}
\newpage
%\printtables
\end{document}